\documentclass[useAMS,usenatbib]{mnras}
\usepackage{graphicx,subfig}
\usepackage{float}
\usepackage{bm}
\usepackage{amsmath}
\usepackage{amsfonts,amssymb}
\usepackage{color}
\definecolor{v}{rgb}{0.6, 0.2, 0.8} 
\definecolor{MAGA}{rgb}{0.1, 0.43, 0.75} 
\usepackage{soul}
\usepackage{enumerate}
\usepackage{float}
\usepackage{ulem}
\usepackage{subfig}
\usepackage{tabularx}
\usepackage{multicol}        
\usepackage{bm}	
\hypersetup{draft}    

\title[Cosmography using SL systems and OHD]{Cosmography using strong lensing systems and cosmic chronometers}

\author[Lizardo, H. Amante, Garc\'{\i}a-Aspeitia, Maga\~na and Motta]{Andr\'es Lizardo$^{1}$ \thanks{E-mail:andres.lizardo@fisica.uaz.edu.mx}, Mario H. Amante$^{1}$\thanks{E-mail:mario.herrera@fisica.uaz.edu.mx}, Miguel A. Garc\'{\i}a-Aspeitia$^{1,2}$\thanks{E-mail:aspeitia@fisica.uaz.edu.mx}, Juan Maga\~na$^{3}$\thanks{E-mail:jmagana@astro.puc.cl}, \newauthor V. Motta$^{4}$\thanks{E-mail:veronica.motta@uv.cl}, \\
$^{1}$ Unidad Acad\'emica de F\'isica, Universidad Aut\'onoma de Zacatecas, \\Calzada Solidaridad esquina con Paseo a la Bufa S/N C.P. 98060,Zacatecas, M\'exico.\\
$^{2}$Consejo Nacional de Ciencia y Tecnolog\'ia, Av. Insurgentes Sur 1582. \\ Colonia Cr\'edito Constructor, Del. Benito Ju\'arez C.P. 03940, Ciudad de M\'exico, M\'exico.\\ $^3$Instituto de Astrof\'isica \& Centro de Astro-Ingenier\'ia, Pontificia Universidad Cat\'olica de Chile, \\Av. Vicu\~na Mackenna, 4860, Santiago, Chile\\
$^4$Instituto de F\'isica y Astronom\'ia, Universidad de Valpara\'iso, Avda. Gran Breta\~na 1111, Valpara\'iso, Chile.  \\}

\date{Accepted YYYYMMDD. Received YYYYMMDD; in original form YYYYMMDD}
\pubyear{????}

\begin{document}
\maketitle

\begin{abstract}
Using a new sub-sample of observed strong gravitational lens systems, for the first time, we present the equation for the angular diameter distance in the $y$-redshift scenario for cosmography and use it to test  the cosmographic parameters. In addition, we also use the observational Hubble data from cosmic chronometers and a Joint analysis of both data is performed. Among the most important conclusions are that this new analysis for cosmography using Strong Lensing Systems is equally competitive to constrain the cosmographic parameters as others presented in literature. Additionally, we present the reconstruction of the effective equation of state inferred from our samples, showing that at $z=0$ those reconstructions from Strong Lensing Systems and Joint analysis are in concordance with the standard model of cosmology.
\end{abstract}

\begin{keywords}
cosmology: theory, cosmological parameters, gravitational lensing: strong.
\end{keywords}

\section{Introduction}

Nowadays it is confirmed that the Universe underwent a transition from a non accelerated to an accelerated evolution at redshift $z\sim0.7$; being the first evidence, the observations obtained from Supernovaes of the Type Ia by different teams \citep[SNeIa,][]{Riess:1998,Perlmutter:1999}. In the framework of general relativity (GR), the cause of this late acceleration is still a mystery, although a plausible explanation is the existence of an entity called the dark energy (DE). Other observational evidence of the Universe acceleration also comes from: Cosmic Microwave Background Radiation \citep[CMB,][]{Aghanim:2018}, Baryonic Acoustic Oscillations \citep[BAO,][]{Eisenstein:2005,Blake:2012,Alam:2017,Bautista:2017}, Hubble parameter measurements \citep{Moresco_2012}, current compilations of SNeIa \citep{Scolnic:2017caz}, among others. In this vein, the most accepted candidate to interpret the DE is the cosmological constant (CC) or also called $\Lambda$, with a measured energy density of $\rho_{\Lambda}^{obs}\leq(10^{-12}\rm{GeV})^4$ \citep{Carroll:2000}. It is commonly assumed that the CC origin comes from the quantum vacuum fluctuations \citep{Zeldovich,Weinberg}, resulting in a theoretical value for the CC energy density of $\rho^{theo}_{\Lambda}\sim(10^{18}$GeV)$^4$, being 120 order of magnitude of difference with the one obtained from observations. Despite its success in describing several observations as an essential component of the $\Lambda$ Cold Dark Matter ($\Lambda$CDM) model, until now it has not been possible to neither reconcile the theoretical and observational value for the energy density nor solve the problem of coincidence, i.e. why the Universe accelerate at late epochs and not at early times?

Confronted with these conundrums of the CC \citep{Carroll:2000,Copeland:2006}, the source of the late cosmic acceleration has been explored using new alternatives like dynamical fields as quintessence and phantom fields \citep{Wetterich:1988,Caldwell:1998,Caldwell:2002,Chiba:1998}, Chaplygin gases \citep{Hernandez-Almada:2018osh}, or the modification of the General Relativity (GR) such as $f(R)$ \citep{Jaime:2012nj}, brane-worlds \citep{Garcia-Aspeitia:2016kak,Garcia-Aspeitia:2018fvw}, unimodular gravity \citep{Perez:2017krv,Garcia-Aspeitia:2019yni,Garcia-Aspeitia:2019yod}, dissipative cosmological models \citep{Cruz:2019wbl}, among others. 
However, these models are theoretical propositions that must be contrasted with diverse cosmological observations before having the last word. From here, a question arises: is it possible to infer the preferred model only through the data of cosmological observations?

In this sense, a model-independent approach to understand the Universe evolution is to demand only the cosmological principle i.e. the homogeneity and isotropy conditions. Hence,
\cite{Chiba:1998} and independently
\cite{Visser:2003} generated some of the first ideas to study the Universe through its kinematic properties; these properties can be related to derivatives of the Hubble parameter for different orders which are dubbed the cosmographic parameters (also known as statefinders parameters). The analysis of the values for these derivatives is the so called Cosmography. Traditionally the cosmography uses Taylor series expanded around the redshift, ($z=0$ more precisely) to approximate the Universe kinematic parameters and the price to pay is that the convergence ratio should be at $z<1$ \citep{Cattoen:2007sk}. One of the ideas to alleviate the convergence problem for $z>1$ is the use of a new parameterization, $y=z/(z+1)$ hereafter called the $y$-redshift, proposed also by \cite{Cattoen:2007sk}. This new form has been employed by \cite{Busti:2015xqa} to estimate the cosmographic parameters ($q_0$, $j_0$, $s_0$ and $l_0$) in the redshift region $0.015<z<1.55$ with SNeIa data from the JLA sample. Another method to avoid the divergence in the Taylor series is to use the Pad\'e approximation \citep{padeaprox}, which is a rational expansion of a function, usually applied to the luminosity distance. Its advantages over the Taylor series is that it can reach those redshift values where the Taylor series diverges. Finally, authors introduce $\arctan(y)$ and $z/(1+z^2)$ functions as a viable parameterizations to study the cosmography  at different stages of the Universe evolution \cite[see][for details]{Aviles:2012ay}.

Recently, \citet{Zhang:2016urt} perform a cosmographic analysis with Taylor series and the $y$-redshift to obtain an expression for the Equation of State (EoS) of dark energy using SNeIa data obtaining that a constant EoS dark energy model may be inappropriate due to the values obtained for the jerk parameter \cite[$j_0 \not= 1$, see also][]{Busti:2015xqa}. They also introduce a novel way to treat the errors associated with the redshift and the $y$-redshift by minimizing a risk function which establishes a balance between the bias and the variance obtained from the observations of SNeIa data. They suggest that a Taylor expansion up to the second order is the best approximation in the SNeIa scenario. In addition, the propagation of errors in cosmography during the statistical analysis has important consequences in the estimation of the cosmographic parameters as stated by \cite{Aviles:2016wel} and extended by \cite{Capozziello:2017nbu} through the use of Chebyshev polynomials.

Until now, several cosmographic estimations have been performed mainly using SNeIa, Gamma Ray Burst \cite[GRB][]{refId0} and Observational Hubble Data (OHD) \citep{Jimenez:2001gg,Magana:2017nfs}. Different cosmological data and techniques are needed to ensure stronger constraints and concordance on the cosmographic parameter values; even studying the cosmography in theories where gravity must be modified \cite[see for example][]{Capozziello:2019cav,TeppaPannia:2018ale}.

On the other hand, strong lensing measurements have been used to analyze DE models since \cite{Grillo:2007iv} proposed a method that relates the mass estimations of the lens galaxy (early type galaxy) using two different measurements: the mass enclosed by the Einstein radius and the mass measured by the central stellar velocity dispersion assuming an isothermal profile ($\sigma^2\propto 1/r^2$). They showed that the central stellar velocity dispersion is comparable enough to the velocity dispersion of the isothermal lens model in the $\Lambda$CDM regime. Since then, many works have analyzed the most competitive DE models using this method. For instance, \citet{Biesiada:2011di} examined the standard model of cosmology $\Lambda$CDM, the $\omega$CDM model and the Chevallier-Polarski-Linder (CPL) parameterization \citep{Chevallier:2000qy,Linder:2002dt}, using a sample with 20 strong lensing measurements from Sloan Lens ACS survey  \citep[SLACS,][]{bolton2006} and Lenses Structure and Dynamics survey \citep[LSD,][]{Treu2004}. Later on, \citet{Cao:2011bg} analyzed the same models using two larger samples of Strong Lensing systems (SLS): the first one with a total of 80 systems, and the second one  with 46 systems containing 36 two-image lenses and 10 lens galaxy clusters with X-ray observations measured by \citet{2011RAA....11..776Y}. These authors showed that the cosmological parameters are sensitive to different strong lensing data, obtaining a non-accelerating Universe for the sample with 80 systems in the CPL model ($\omega_0=0.6$); however being consistent at 1$\sigma$ with an accelerated one. Other dark energy models, like Dvali-Gabadadze-Porrati (DGP) and Ricci model along with the $\Lambda$CDM and $\omega$CDM models, were studied by \cite{Cao:2012ja} using 122 SLS. They support an accelerated expansion of the Universe according to the standard model of cosmology ($\Lambda$CDM), but without finding significant evidence against the other models \cite[see also][]{Cao:2015qja}. Recently, \citet{Amante:2019xao} compiled the largest strong lensing sample, produced by single (early type) lens galaxies, which contains 204 points with  $0.0625<z_l<0.958$ for the lens and $0.196<z_s<3.595$ for the source. The authors constrain three DE models,  in different regions of the data, demonstrating that these data are able to constraint the DE parameters. \citep[see also ][for recent compilations regarding strong lensing measurements] {Leaf:2018lfu,Chen:2018jcf}. Moreover, the authors present a strong lensing system fiducial sample containing $143$ data points which can provide better constraints on DE parameters.

In this work, we use such fiducial subsample of 143 SLS of the full compilation by \citet{Amante:2019xao} to test the first four cosmographic parameters: ($q_{0}$, $j_{0}$, $s_{0}$, and $l_{0}$) with the $y$-redshift parameterization Taylor expansion approach. The Einstein radius and velocity dispersion for each SLS was measured by different surveys: Sloan Lens ACS survey \citep{bolton2006}; CfA-Arizona Space Telescope LEns Survey \citep{Munoz:1999qe}; BOSS Emission-Line Lens Survey \citep{More_2012}; Lenses Structure and Dynamics survey \citep{Koopmans_2004}; CFHT Strong-lensing Insight into Dark Energy Survey \citep{Treu_2018}. Moreover, we will use OHD compiled by \cite{Magana:2017nfs} as an additional data sample to constraint the same parameters. Finally, using the equations in terms of these parameters, we present the results for the EoS under the three (SLS, OHD and Joint) mentioned samples.

The outline of the paper is as follows. Sec. \ref{MP} discuss the mathematical details to construct the angular diameter distance ratio of SLS analysis under the cosmographic approach. In Sec. \ref{data} we present the data and methodology to estimate the cosmographic parameters. In Sec. \ref{Res} we show the results and finally in Sec. \ref{Con} we remark the conclusions and outlooks.

\section{Mathematical Preliminaries} \label{MP}

The fundamental attributes of the evolution of the Universe can be either kinematic or dynamic, with the kinematic characteristics independent of any cosmological model unlike the dynamical ones.
In this vein, the analysis of the kinematic characteristics of the Universe based on the cosmological principle is encompassed by the cosmography. 

We start this analysis by introducing the first five cosmographic parameters in terms of the scale factor $a(t)$, 
\begin{eqnarray}
    H(t) &=& + \frac{1}{a} \frac{da}{dt}, \\
    q(t) &=& - \frac{1}{a} \frac{d^2a}{dt^2} \left[\frac{1}{a} \frac{da}{dt}\right]^{-2}, \label{q} \\
   j(t) &=& + \frac{1}{a} \frac{d^3a}{dt^3} \left[\frac{1}{a} \frac{da}{dt}\right]^{-3}, \label{j} \\
    s(t) &=& + \frac{1}{a} \frac{d^4a}{dt^4} \left[\frac{1}{a} \frac{da}{dt}\right]^{-4}, \label{s} \\
    l(t) &=& + \frac{1}{a} \frac{d^5a}{dt^5} \left[\frac{1}{a} \frac{da}{dt}\right]^{-5}, \label{l}
\end{eqnarray}
with the Hubble, deceleration, jerk, snap and lerk parameters, respectively \citep{Demianski:2012}. It is worth to notice that $H(t)$ is related to the expansion rate of the Universe, $q(t)$ is associated with the acceleration/deceleration rate, and $j(t)$ indicates whether DE is a constant function of time or not, in particular, the case $j=1$ specifies a CC. The $s(t)$ and $l(t)$ parameters are corrective to high order in the Taylor expansion, but their physical interpretation is not clear.
It is important to mention that this set of cosmographic parameters provide a method to analyze models by comparing their values with those obtained within the framework of each model.

\subsection{Cosmography in z-redshift}

With the parameters displayed in equations \eqref{q} to \eqref{l} we can rewrite the Hubble parameter expanded as a Taylor series around $z=0$, restricted to the spatially flat case ($k=0$) as

\begin{eqnarray}
    \nonumber H(z) & = & H_0 + \frac{dH}{dz}\Big|_{z=0}z + \frac{1}{2!}\frac{d^2H}{dz^2}\Big|_{z=0}z^2 
    \nonumber \\
     &+&
    \frac{1}{3!}\frac{d^3H}{dz^3}\Big|_{z=0}z^3 + \dots \nonumber \\
     & = & H_0 \Big[ 1 + (1+q_0)z +\frac{1}{2}(-q_0^2 + j_0)z^2
     \nonumber \\
     &+& \frac{1}{6}(3q_0^2 + 3q_0^3 -4q_0j_0 - 3j_0 -s_0)z^3 
     \nonumber \\
     &+& \frac{1}{24}(-12q_0^2 - 24q_0^3 - 15q_0^4 + 32q_0j_0 + 25q_0^2j_0
     \nonumber \\
     &+& 7q_0s_0 + 12j_0 - 4j_0^2 + 8s_0 + l_0)z^4 \nonumber \\ & +&\mathcal{O}(z^5) \Big]. \label{Hubblez}
\end{eqnarray} 
Here the subscript $0$ means that they are evaluated at the present time ($z=0$).
Hence, the luminosity distance \citep{Zhang:2016urt} is written in the form
\begin{equation}
    d_L(z) = d_H (z + \mathcal{K}_1 z^2 + \mathcal{K}_2 z^3 + \mathcal{K}_3 z^4 + \mathcal{K}_4 z^5), \label{dl}
\end{equation}
where $d_H = c/H_0$ and the coefficients are
\begin{eqnarray}
    \mathcal{K}_1 &=& \frac{1}{2} (1-q_0), \\
    \mathcal{K}_2 &=& - \frac{1}{6} (1-q_0-3q_0^2+j_0), \\
    \mathcal{K}_3 &=& \frac{1}{24} (2 - 2q_0 - 15q_0^2 - 15q_0^3 + 5j_0 + 10q_0j_0 \nonumber\\&& +s_0), \\
    \mathcal{K}_4 & = & \frac{1}{120} (-6 + 6q_0 + 81q_0^2 + 165q_0^3 + 105q_0^4
    \\ \nonumber
    & \ & + 10j_0^2 - 27j_0 - 110q_0j_0 - 105q_0^2j_0 
    \\ \nonumber
    & \ & - 15q_0s_0 - 11s_0 - l_0).
\end{eqnarray}
Following the guidelines of \cite{Demianski:2012}, we define the physical distance as the distance traveled by a photon that is emitted at time $t_*$ and absorbed at the current epoch $t_0$ which is
\begin{equation}
    d_Z = c \int dt = c(t_0-t_*),
\end{equation} 
hence, we can construct the series for $z(d_Z)$
\begin{equation}
    z(d_Z) = p +\mathcal{Z}_d^1p^2+\mathcal{Z}_d^2p^3+\mathcal{Z}_d^3p^4+\mathcal{Z}_d^4p^5
\end{equation}
where $p =H_0d_Z/c$ and
\begin{eqnarray}
    \mathcal{Z}_d^1 &=& 1+\frac{q_0}{2}, \\
    \mathcal{Z}_d^2 &=& 1 + q_0 + \frac{j_0}{6}, \\
    \mathcal{Z}_d^3 &=& 1 + \frac{3}{2}q_0+\frac{q_0^2}{4}+\frac{j_0}{3}-\frac{s_0}{24}, \\
    \mathcal{Z}_d^4 &=& 1 + 2q_0 + \frac{3}{4}q_0^2 + \frac{q_0j_0}{6} + \frac{j_0}{2}-\frac{s_0}{12}+l_0.
\end{eqnarray}
and to obtain the physical distance expressed as a function of redshift we reverse the series $z(d_Z) \to d_Z(z)$.

\begin{equation}
    d_Z(z) = d_H(z + Q_1 z^2 + Q_2 z^3 + Q_3 z^4 + Q_4 z^5), \label{dz}
\end{equation}
where the coefficients are 
\begin{eqnarray}
    Q_1 &=& -\left(1+\frac{q_0}{2}\right), \\
    Q_2 &=& 1 + q_0 + \frac{q_0^2}{2} - \frac{j_0}{6}, \\
    Q_3 &=& -1 - \frac{3}{2}q_0 - \frac{3}{2}q_0^2 - \frac{5}{8}q_0^3 + \frac{1}{2}j_0 + \frac{5}{12}q_0j_0 \nonumber\\ &&+ \frac{1}{24}s_0, \\
    Q_4 & = & 1 + 2q_0 + 3q_0^2 + \frac{5}{2}q_0^3 + \frac{7}{2}q_0^4 - \frac{5}{3}q_0j_0 - \frac{7}{8}q_0^2j_0 \nonumber \\
    &&-\frac{1}{8}q_0s_0 - j_0 + \frac{j_0^2}{12} - \frac{s_0}{6} - \frac{l_0}{120}.
\end{eqnarray}
As is indicated by \cite{Hogg:1999ad}, the transverse comoving distance, $\Delta$, between two nearby objects in a flat universe is related to the line of sight comoving distance, which is written in terms of the $d_Z$ as
\begin{equation}
    \Delta(d_Z) = \frac{d_Z(z)}{a_0} \left[1 + \mathcal{R}_1 S + \mathcal{R}_2 S^2 + \mathcal{R}_3 S^3 + \mathcal{R}_4 S^4 + \mathcal{R}_5 S^5 \right], \label{DZ}
\end{equation}
where $a_0$ is the scale factor at $z=0$ and it will be fixed as $a_0=1$ for convenience, with $S=H_0 d_Z(z)/c$ and the coefficients are
\begin{eqnarray}
    \mathcal{R}_1 &=& \frac{1}{2}, \\
    \mathcal{R}_2 &=& \frac{1}{6}(2+q_0), \\
    \mathcal{R}_3 &=& \frac{1}{24}(6 + 6q_0 + j_0), \\
    \mathcal{R}_4 &=& \frac{1}{120}(24 + 36q_0 + 6q_0^2 + 8j_0 - s_0), \\
    \mathcal{R}_5 &=& \frac{1}{144}(24 + 48q_0 + 18q_0^2 + 4q_0j_0 + 12j_0 - 2s_0\nonumber\\&& + 24l_0).
\end{eqnarray}
Notice that these equations are valid at $z<1$ to avoid convergence problems.

\subsection{Cosmography in $y$-redshift}

As we mention before, to solve this problem  we introduce a reparameterization \citep{Cattoen:2007sk}, called the \textit{y-redshift}, as
\begin{equation}
    y = \frac{z}{1+z},
    \label{y}
\end{equation}
with this parameterization the limits in our analysis transforms as: $z \in \left[0,\infty\right)$, $y\in\left[0,1\right]$ into $z \in \left[-1,0\right]$ and  $y\in\left(-\infty,0\right]$. Therefore, it is possible to add this parameterization to the expressions of the Hubble parameter as \citep{Aviles:2012ay}\footnote{This equation is corrected for the sign error in \cite{Aviles:2012ay} paper.}
\begin{eqnarray}
    H(y) & = &  H_0 \Big[ 1 + (1+q_0)y +\frac{1}{2}(2+2q_0-q_0^2 + j_0)y^2 \nonumber \\
     &+& \frac{1}{6}(6+6q_0-3q_0^2 + 3q_0^3 -4q_0j_0 + 3j_0 -s_0)y^3 \nonumber \\
     &+& \frac{1}{24}(24-4j_0^2+l_0+24q_0-12q_0^2+12q_0^3-15q_0^4
     \nonumber\\
     &+& j_0(12-16q_0+25q_0^2)-4s_0+7q_0s_0)y^4 \nonumber \\  &+&\mathcal{O}(y^5) \Big], \label{Hubbley}
\end{eqnarray}
The luminosity distance in terms of the \textit{y-redshift} results
\begin{equation}
    d_L(y) =  d_H(y + \mathcal{C}_1 y^2 + \mathcal{C}_2 y^3 + \mathcal{C}_3 y^4 + \mathcal{C}_4 y^5),
    \label{DLy}
\end{equation}
being

\begin{eqnarray}
    \mathcal{C}_1 &=& \frac{1}{2} (3-q_0), \\
    \mathcal{C}_2 &=& \frac{1}{6} (11-5q_0-3q_0^2-j_0), \\
    \mathcal{C}_3 &=& \frac{1}{24} (50 - 26q_0 + 21q_0^2 - 15q_0^3 - 7j_0 + 10q_0j_0 \nonumber\\&& + s_0), \\
    \mathcal{C}_4 & = & \frac{1}{120} (274 - 154q_0 + 141q_0^2 - 135q_0^3 + 105q_0^4
    \\ \nonumber
    & \ & + 10j_0^2 - 47j_0 + 90q_0j_0 - 105q_0^2j_0 
    \\ \nonumber
    & \ & - 15q_0s_0 + 9s_0 - l_0).
\end{eqnarray}
The physical distance becomes

\begin{equation}
    d_Z(y) = d_H(y + \mathcal{Q}_1 y^2 + \mathcal{Q}_2 y^3 + \mathcal{Q}_3 y^4 + \mathcal{Q}_4 y^5), \label{dzy}
\end{equation}
where
\begin{eqnarray}
    \mathcal{Q}_1 &=& -\frac{1}{2}q_0, \\
    \mathcal{Q}_2 &=& \frac{1}{2}q_0^2 - \frac{1}{6}j_0, \\
    \mathcal{Q}_3 &=& -\frac{5}{8}q_0^3 + \frac{1}{3}j_0 + \frac{5}{12}q_0j_0 + \frac{1}{24}s_0, \\
    \mathcal{Q}_4 &=& \frac{7}{2}q_0^4 - \frac{7}{8}q_0^2j_0 + \frac{1}{12}j_0^2 - \frac{1}{8}q_0s_0 - \frac{1}{120}l_0.
\end{eqnarray}
Finally, the transverse comoving distance takes the form
\begin{equation}
    \Delta(d_Z) = \frac{d_Z(y)}{a_0} \left[1 + \mathcal{R}_1 S + \mathcal{R}_2 S^2 + \mathcal{R}_3 S^3 + \mathcal{R}_4 S^4 + \mathcal{R}_5 S^5 \right], \label{DZ-y}
\end{equation}
where $S=H_0d_Z(y)/c$ and $\mathcal{R}_1$ to $\mathcal{R}_5$ are the same as those shown in Eqs. (18)-(22). Notice that $d_L(y)$ is constructed with the luminosity distance but now in terms of the $y$-redshift and $a_0=1$ for convenience.

\section{Data and Methodology} \label{data}

\subsection{Strong Lensing} \label{SL}

As we mentioned earlier, the aim of this paper is to tackle the cosmography analysis with strong lensing measurements using the expansion for the angular-diameter distance. 
According to \citet{Schneider_book:1992}, when a lensing galaxy is modeled as a Singular Isothermal Sphere (SIS) the Einstein radius is
\begin{equation}
\theta_{E} = 4 \pi \frac{\sigma^2_{SIS} D_{ls}}{c^2 D_{s}}, \label{thetaESIS}
\end{equation}
where $\sigma_{\mathrm{SIS}}$ is the velocity dispersion of the lensing galaxy, $D_{ls}$ is the angular diameter distance between the lens and source, $c$ is the speed of light, and $D_{s}$ is the angular diameter distance to the source. We can express the ratio between the angular diameter distances as
\begin{equation}
    D^i \equiv D_{ls}/D_{s}, \label{dth}
\end{equation}
where the superscript $i$, means that it can be applied to either observational or theoretical angular diameter distances.
Thus, from the equation for the Einstein radius 
\eqref{thetaESIS} we can construct an observational quantity, a ratio between angular diameter distances, as
\begin{equation}
D^{obs}\equiv\frac{D_{ls}}{D_{s}} = \frac{c^2 \theta_{E}}{4 \pi \sigma^2}, \label{Dlens}
\end{equation} 
where $\sigma$ is the measured spectroscopic velocity dispersion of the lens dark matter halo. A corrective parameter $f$ can be introduced in Eq. \eqref{Dlens} to take into account possible systematic differences among systems \citep[e.g. elliptical instead of spherical profile for the lens halo, line-of-sight stellar velocity dispersion as opposed to the dark matter halo  velocity dispersion, see for example][]{Ofek:2003sp,Cao:2011bg}. \citep{Treu:2006ApJ,Amante:2019xao}. In Appendix \ref{AP2} we will discuss the effects of including this factor in the estimation of cosmographic parameters.

The relation between the luminosity distance and the angular-diameter distance, $D(z)$, reads as \citep{Hogg:1999ad} for a flat Universe

\begin{equation}
    D(z) = \frac{d_L(z)}{(1+z)^2},
\end{equation}
where $d_L(z)$ is the luminosity distance given by equation (\ref{dl}). In addition, using the parameterization shown in Eq.  \eqref{y}, we propose:

\begin{equation}
    D(y) = (1-y)^2 d_L(y),
\end{equation}
expanding the previous relation in a Taylor series, we finally obtain:
\begin{eqnarray}
&D(y)&=d_H[y+ (\mathcal{C}_1 - 2)y^2 + (1-2\mathcal{C}_1+\mathcal{C}_2)y^3+ \nonumber\\
    && (\mathcal{C}_1 -2\mathcal{C}_2 + \mathcal{C}_3)y^4 + (\mathcal{C}_2 - 2\mathcal{C}_3 + \mathcal{C}_4)y^5], \label{Dy}
\end{eqnarray}
which is written in terms of the coefficients of Eq. (\ref{DLy}).\\

To estimate the cosmographic parameters we need a theoretical expression for $D_{ls}$ in Eq. \eqref{dth}, i.e. the angular diameter distance between two objects at different redshifts as presented by \cite{Hogg:1999ad}, which in our case is
\begin{equation}\label{dthz}
    D_{ls}(z) = \left(\frac{1}{1+z_s}\right)\left[\Delta_s - \Delta_l \right],
\end{equation}
then the ratio $D_{ls}/D_s$ for $z$-redshift takes the form
\begin{equation}
\frac{D_{ls}}{D_{s}} \equiv D^{th}\left(z_{l}, z_{s}; \bf{\Theta} \right) = \frac{\left(1+z_s\right)^{-1} \big[\Delta(z_s) - \Delta(z_l)\big]}{D(z_s)}. \label{einstradz}
\end{equation}

On the other hand, Eq. \eqref{dthz} for the $y$-parameterization \eqref{y} reads as
\begin{equation}\label{dthy}
    D_{ls}(y) = (1-y_s)\left[\Delta_s - \Delta_l \right],
\end{equation}
being $\Delta_l$ and $\Delta_s$ the comoving distances to the lens and source in both \eqref{dthz} and \eqref{dthy}, respectively; under the constraint $0<y_s<1$. Thus, Eq. \eqref{dth} results
\begin{equation}
\frac{D_{ls}}{D_{s}} \equiv D^{th}\left(y_{l}, y_{s}; \bf{\Theta} \right) = \frac{ (1-y_s) \big[\Delta(y_s) - \Delta(y_l)\big]}{D(y_s)}, \label{einstrad}
\end{equation}
where $y_{l}$, $y_s$ are the $y$-redshifts evaluated at $z=z_l$ and $z=z_s$ respectively and  $\mathbf{\Theta}$ is the vector containing the cosmographic parameters. 
It is worth to mention that the denominator of this previous equation is an expression of the angular diameter distance written in Eq. \eqref{Dy}, and the numerator stands for Eq. \eqref{dthy}.
Regarding the convergence problems of considering just the $z$-redshift without limits, we will focus our analysis with Eq. \eqref{einstrad}.

To analyze the cosmographic parameters we use the most recent compilation of SLS given by \citet{Amante:2019xao} which consist of 204 systems spanning the redshift region $0.065<z_l<0.96$ for the lens and $0.196<z_s<3.6$ for the source. This sample considers only those systems with elliptical galaxies as lenses and known spectroscopic velocity dispersion, lens and source redshifts ($z_l$,  $z_s$), and the Einstein  radius $\theta_E$. In this work we only use SLS with an observational lens equation ($D^{obs}$) within the region $ 0.5 \leq D^{obs} \leq 1$. This region is chosen because it allows  better cosmological constraints by avoiding non-physical regions ($ D^{obs} > 1$) or those ($ D^{obs} < 0.5$) where the theoretical lens equation \eqref{einstrad} can not be accurately modeled for some measured systems, yielding a non-accelerating Universe at late times for the cosmological parameter constraints \cite[see][for further details]{Amante:2019xao}. 
With all these considerations we obtain a sample of 143 SLS between the range of $0.0625 < z_l < 0.958 $ for  the lens galaxy and $0.2172 < z_s < 3.595 $ for the source. 
Notice that using the parameterization in Eq. \eqref{y} the frontier values for the $y$-redshift of the lens are $[0.058, 0.489]$ and for the source $[0.163, 0.782]$. For a further analysis we take into account a set of simulated SLS in order to check if we can recover the cosmographic parameters considering an aleatory sample of data (see Appendix \ref{AP3}).

Thus, the cosmographic parameters can be contrasted by minimizing the following chi-square function,
\begin{equation}\label{chi}
\chi_{\mbox{SLS}}^2 = \sum_{i=1}^{N_{SLS}} \frac{ \left[ D^{th}\left(y_{l}, y_{s}; \bf{\Theta} \right)  -D^{obs}(\theta_{E},\sigma^2)\right]^2 }{ (\delta D^{\rm{obs}})^2},
\end{equation}
where the sum is over all the $N_{SLS}=143$, lens systems and $\delta D^{\rm{obs}}$ is the uncertainty of  each $D^{obs}$ measurement computed in the standard way of error propagation as 
\begin{equation}
\delta D^{\rm{obs}}= D^{\rm{obs}} \sqrt{\left( \frac{\delta \theta_E}{\theta_E} \right)^2 + 4 \left( \frac{\delta \sigma}{\sigma} \right)^2},
\end{equation}
being $\delta \theta_E$ and $\delta \sigma$ the errors reported for the Einstein radius and velocity dispersion, respectively. For those systems where $\delta \theta_E$ is not reported, following the guidelines of \cite{Cao:2015qja}, we assume an error of $\delta \theta_E=0.05\arcsec$, which is the average value of the systems with reported errors in the sample \citep{Amante:2019xao}.

\subsection{Observational Hubble Data} \label{CC}

Another test to consider is the data from the so-called Observational Hubble Data (OHD). First suggested by \cite{Jimenez:2001gg}, the method uses the differential age between pairs of passive evolving galaxies in the zone of $0.07 < z < 1.97$ with similar metallicity and separated by a small redshift interval \citep[for instance,][measure $\Delta z\sim0.04$ at $z<0.4$ and $\Delta z\sim0.3$ at $z>0.4$]{Moresco_2012}.
Thus, the expansion rate is written as
\begin{equation}
    H(z) = - \frac{1}{1+z} \frac{dz}{dt}, \label{Loeb}
\end{equation}
where $dz$ can be measured with high accuracy 
\citep[spectroscopic redshifts of galaxies have typical uncertainties $\sigma_z < 0.001$,][]{Moresco_2012}.
Notice that Eq. \eqref{Loeb} can be written in terms of the y-redshift as
\begin{equation}
    H(y) = - \frac{1}{1-y} \frac{dy}{dt}. \label{Loeb-y}
\end{equation}
With a set of 31 systems of OHD compiled by \cite{Magana:2017nfs}, we can estimate the cosmographic parameters using the Taylor expansion of the Hubble law in terms of the last ones and the $H(y)$ values by applying a $\chi^2$ test 
\begin{equation}
    \chi^2_{OHD} = \sum_{i=1}^{N_{OHD}} \frac{ \left[ H^{th}(y)  -H^{obs}(y)\right]^2 }{ \sigma^{2}_{H}}, \label{chiCC}
\end{equation}
where $H^{th}$ is the theoretical Hubble parameter in terms of the cosmographic parameters and the y-redshift, $H^{obs}$ is the observational one, and $\sigma_{H}$ its uncertainty. 

\subsection{Joint analysis}

We also perform a joint analysis of the SLS and cosmic chronometers data to estimate the cosmographic parameters. The $\chi^2$ function for this joint analysis is
\begin{equation}
    \chi^2_{joint}=\chi^2_{SLS} + \chi^2_{OHD}, 
    \label{eq:chijoint}
\end{equation}
where $\chi^2_{SLS}$ and $\chi^2_{OHD}$ are given by Eqs. \eqref{chi} and \eqref{chiCC}  respectively. Thus, the joint analysis can improve the constraints on the cosmographic parameter by breaking the degeneracy. 

\subsection{The construction of the effective EoS} \label{EoS}

To find an effective EoS in $y$-redshift cosmography \footnote{See \cite{Aviles:2012ay} for the EoS in terms of the redshift parameter.} we construct the relation $w_{eff}(y)=P(y)/\rho(y)$, where the pressure $P$ is
\begin{equation}
    P(y)=\sum_{k=0}^{\infty}\frac{1}{k!}\frac{d^kP}{dy^k}\Big\vert_{y=0}y^k, \label{pressureP}
\end{equation}
where
\begin{eqnarray}
    P(y=0)&=&H_0^2(2q_0-1), \\
    \frac{dP}{dy}\Big\vert_{y=0}&=&2H_0^2(j_0-1), \\\frac{d^2P}{dy^2}\Big\vert_{y=0}&=&H_0^2(-1-3q_0+j_0-q_0j_0-s_0), \\
    \frac{d^3P}{dy^3}\Big\vert_{y=0}&=&\frac{H_0^2}{3}[-6-j^2_0+l_0-6q_0+3q^2_0+j_0(-3\nonumber\\&&-2q_0+3q_0^2)-2s_0+3q_0s_0].
\end{eqnarray}
The density is related to the Hubble parameter by $\rho(y)=3H^2(y)$, where the Hubble parameter as a function of $y$-redshift is given by Eq. \eqref{Hubbley}. Notice that, for $y=0$, the effective equation of state will be
\begin{equation}
    w_{eff}(0)=\frac{1}{3}(2q_0-1),
\end{equation}
which only depends of $q_0$, and the other cosmographic parameters will only play a role if $y\neq0$ ($z\neq0$).

\section{Results} \label{Res}

We start this section presenting the $\Lambda$CDM cosmographic parameters, which hereafter will be used as reference for the priors in our analysis. Following \citet{Dunsby:2015ers,Bolotin:2018xtq}, we have
\begin{eqnarray}\label{qq}
    q_0 &=& -1 + \frac{3}{2} \Omega_{0m}, \\
    j_0 &=& 1, \\
    s_0 &=& 1 - \frac{9}{2}\Omega_{0m}, \\
    l_0 &=& 1 + 3\Omega_{0m} + \frac{9}{2}\Omega_{0m}^2, \label{ll}
\end{eqnarray}
where the subscript 0 indicates the value at $z=0$. Assuming $\Omega_{0m}=0.311$ \citep{Aghanim:2018}, those values are $q_0=-0.53$, $j_0=1$, $s_0=-0.39$ and $l_0=2.36$.
As it is explained in the work of \cite{Dunsby:2015ers}, choosing wide priors for our fits might affect our results producing large errors in them. To avoid this, we followed their guidelines and choose the priors as it is shown in Table \ref{tab:priors}. If numerical outcomes provide good results with small errors, then one can assume the priors are correctly defined. In addition, the $H_0$ parameter was fixed to the value provided by \cite{Riess:2016jrr}, because it is based in near observations, therefore, $H_0= 73.24 \pm 1.79$ $\rm km \, s^{-1} \, Mpc^{-1}$.

As we mentioned previously, we can constrain the cosmographic parameters by minimizing the chi-square functions: for SLS (Eq. \ref{chi}), OHD (Eq. \ref{chiCC}), and the joint analysis (Eq. \ref{eq:chijoint}). We use the \texttt{emcee} Python code \citep{Foreman_Mackey_2013} with 5000 steps on the Markov Chain Monte Carlo (MCMC) phase, a set of 1000 steps for the burn-in phase and $1000$ walkers, having convergence for each of the free parameters in all cases. We performed the analysis for different orders in the expansion of the Taylor series for the luminosity and comoving distances and found that, up to the fourth order, these are good approximations to the Hubble parameter. However, as we are analyzing the parameters on a present time $(z=y=0)$, which is the extreme of the interval of expansion of the Taylor series, the approximations show larger deviations on the other side of the interval of expansion $(y \sim 1)$. 

We discard the SLS that do not satisfy the condition $y_s\leq0.6$. For those systems the approximation for the angular diameter distance of the source ($D_{s}$) fails, i.e. they show values larger than those obtained from computing the angular diameter distances through integrals \citep[see for instance][]{Amante:2019xao}. In addition, we found that the SLS beyond $y_s > 0.6$ give non-physical values in the theoretical lens equation \eqref{einstrad}, i.e. $D^{th}>1$, when the $\Lambda$CDM cosmographic parameters are employed. Hereafter, we discarded 44 SLS with $y_{s}>0.6$ and we just admit the ones with $y_s\leq0.6$ ($99$ SLS). Hence, the following results are obtained under the previously mentioned constraint. For completeness, Appendix \ref{AP1} shows the analysis without the constriction $y_s\leq0.6$. It is worth to notice that this behaviour is only seen for the $y$-redshift parameterization, and different criteria should be used in other parametrizations.

\begin{table}
    \centering
    \begin{tabular}{|c|c|}
    \hline
        Parameter &  Prior \\
    \hline
        $q_{0}$  & Flat in $[-0.95,-0.3]$  \\
    \hline
        $j_{0}$ & Flat in $[0,2]$  \\
    \hline
        $s_0$ & Flat in $[-2,7]$  \\
    \hline
        $l_0$ & Flat in $[-5,10]$  \\
    \hline
    \end{tabular}
    \caption{Priors used for the cosmographic parameters in the MCMC analysis.} 
    \label{tab:priors}
\end{table}

\begin{figure}
    \includegraphics[width=0.5\textwidth]{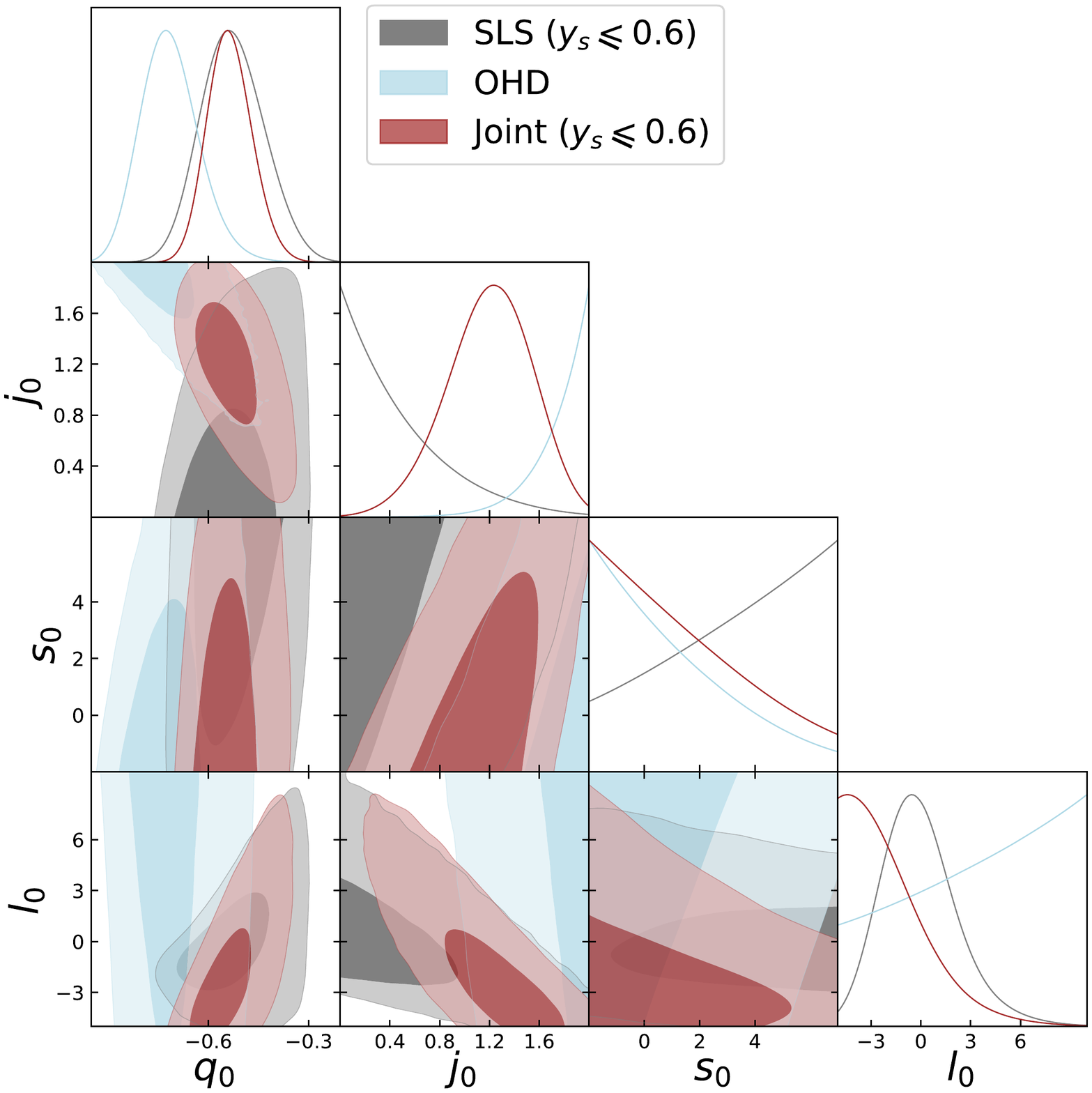}
    \caption{1D marginalized posterior distributions and the 2D $68\%$, $99.7\%$ confidence levels for $q_0$, $j_0$, $s_0$ and $l_0$ parameters from SLS under the constriction $y_s\leq0.6$, OHD and a Joint analysis. The series were truncated up to the fourth order.}
    \label{q0j0s0l0ys<}
\end{figure}

Figure \ref{q0j0s0l0ys<} shows the results of our MCMC analysis for the free parameters, with 2D contours at $68\%$ ($1\sigma$) and $99.7\%$ ($3\sigma$) confidence level (CL), and their corresponding 1D posterior distributions where the series is truncated up to fourth order, under the constriction $y_s\leq0.6$. Our analysis shows that  the values estimated in the joint analysis are closer to the values expected in the standard model ($\Lambda$CDM), using the fourth order approximation.

Table \ref{SLSandCronomvalues} shows the mean values of the cosmographic parameters and their uncertainties within $1\sigma$ of confidence level (CL) for the SLS, OHD, and joint analysis respectively. We also give the minimum chi-square, $\chi_{min}$, and the reduced $\chi_{red}=\chi_{min}/{\rm d.o.f.}$, where the degree of freedom (d.o.f.) is the difference between the number of data points and the free parameters. For the OHD case, the parameter $j_0$ is barely above $3\sigma$ level estimated with the expected values from the $\Lambda$CDM model. In all the other cases, including the SLS and the Joint analysis, it can be argued that we are into the $3\sigma$ frontier. Thus, we use these two sets of data close to the $3\sigma$ region of $\Lambda$CDM to estimate the cosmographic parameters. It is also worth to mention that, after the mock analysis with the $0.5\leq D^{obs}\leq1.0$ data, we obtain consistency for the $j_0$ parameter at $1\sigma$ level with those expected for the $\Lambda$CDM model (see Appendix \ref{AP3}).

Notice that, although in general SLS data is more restrictive\footnote{The values for the cosmographic parameters have smaller error bars.} than SNeIa data presented by \cite{Zhang:2016urt}\footnote{The authors report $-0.74\pm0.45$, $6.16\pm10.06$, $89.06\pm160.80$ and $2056.97\pm2478.44$ for $q_0$, $j_0$, $s_0$ and $l_0$ respectively.}, they produce tighter constraints for $s_0$ and $j_0$. When SLS and cosmic chronometers are considered together (joint analysis), they generate tighter constraints than for SLS alone, in concordance with previous estimations of the cosmographic parameters \cite[see][]{Busti:2015xqa}. 
However, in a recent study, \cite{Rezaei:2020lfy} use a sample of SNeIa (Pantheon), GRB and Quasars to obtain the values $-0.819\pm0.065$, $2.21^{+0.37}_{-0.42}$, $-3.44^{+0.46}_{-1.5}$ and $-3.8^{+8.2}_{-6.2}$ for the four cosmographic parameters, with an effective EoS of $-1.42^{+0.15}_{-0.13}$. Despite that our results differ from those from \cite{Rezaei:2020lfy}, we obtain a better concordance with those expected from $\Lambda$CDM, validating the effectiveness of adding the SLS sample in a Joint analysis. 

\begin{figure}
  \includegraphics[width=0.5\textwidth]{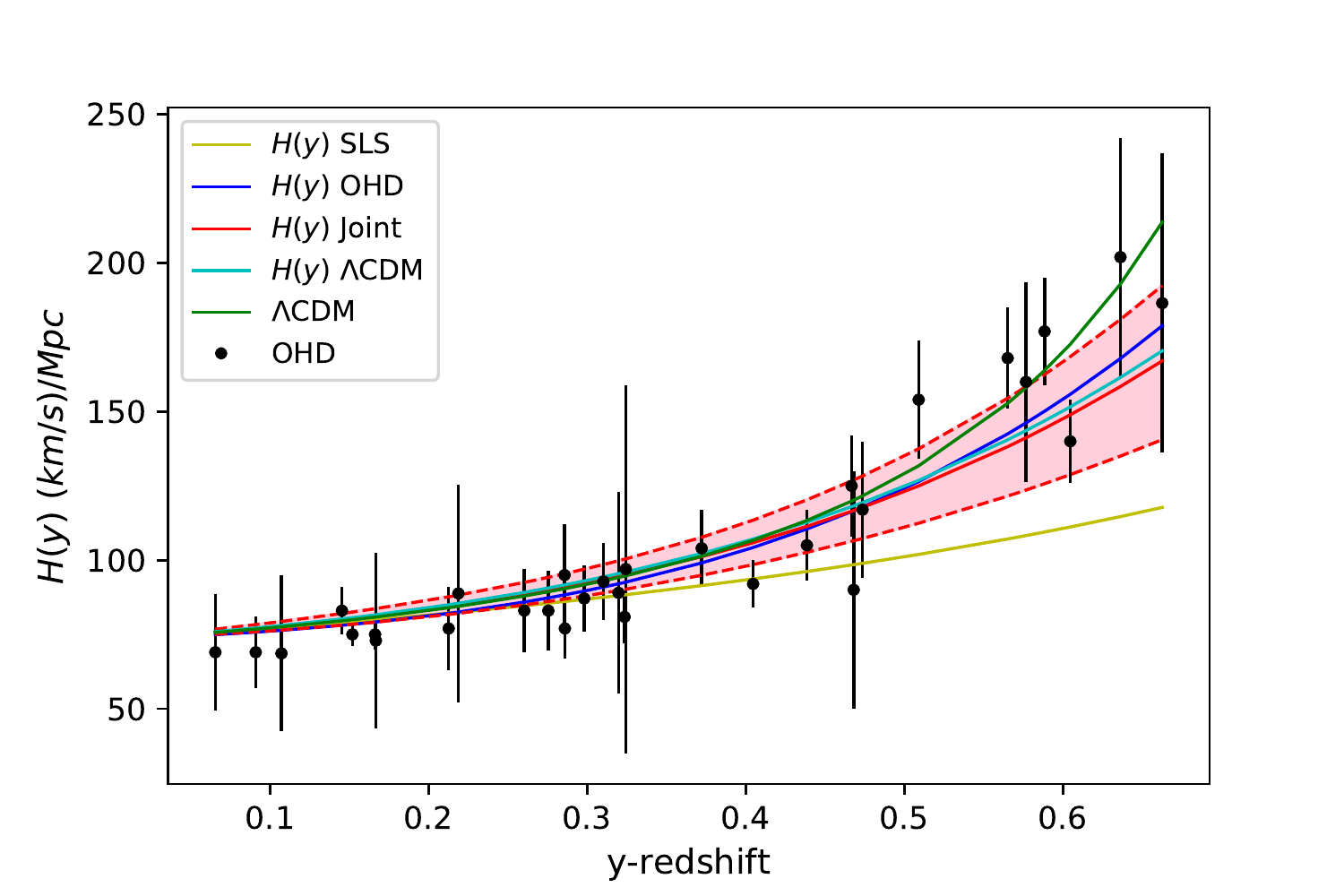} 
\caption{The $H(y)$ in terms of the cosmographic parameters (Eq. \ref{Hubbley}) using the values for each of the analysis performed (SLS, OHD and Joint). We show also the $H(y)$ (Eq. \ref{Hubbley}) with the values expected for the cosmographic parameters given by Eq. \eqref{qq}-\eqref{ll}. The green line corresponds to the traditional Friedmann equation for $\Lambda$CDM model with $\Omega_{0m}=0.311$ \citet{Aghanim:2018}. The shadow region between the dotted red lines corresponds to the $3\sigma$ error for the Joint analysis. The observational Hubble data are shown in the black dots for comparison.} 
\label{fig:HJoint}
\end{figure}

Figure \ref{fig:HJoint} shows the $H(y)$ curve (eq. \eqref{Hubbley}) reconstructed with the mean values for the parameters estimated from each of the analysis presented in the Table \ref{SLSandCronomvalues} together with the one reconstructed using the theoretical values for the $\Lambda$CDM model (eq. \eqref{qq} to \eqref{ll}). The OHD data-set is included for comparison. This plot shows that the Hubble function constructed from cosmographic parameters using the Joint analysis, agrees with the consensus model. Our construction of the $H(y)$ in Taylor series with the parameter values obtained from the Joint and OHD analysis are in good agreement with the observations within the errors; not so the construction of the $H(y)$ with the values obtained from the SLS data.

\begin{table}
    \centering
    \begin{tabular}{|c|c|c|c|c|}
    \hline
        Param. & 99 SLS & 31 OHD & Joint & $\Lambda$CDM  \\
    \hline
    \multicolumn{5}{|c|}{}\\
    \hline
        $q_0$ & $-0.54^{+0.10}_{-0.08}$ & $-0.72^{+0.09}_{-0.07}$ & $-0.54^{+0.07}_{-0.06}$ & $-0.53$\\
    \hline
        $j_0$ & $0.35^{+0.52}_{-0.26}$ & $1.82^{+0.13}_{-0.27}$ & $1.22^{+0.29}_{-0.34}$ & $1$ \\
    \hline
        $s_0$ & $3.70^{+2.36}_{-3.31}$ & $0.44^{+2.98}_{-1.77}$ & $0.86^{+3.13}_{-2.05}$ & $-0.39$\\
    \hline
        $l_0$ & $-0.35^{+2.22}_{-1.60}$ & $4.00^{+4.28}_{-5.61}$ & $-2.31^{+2.74}_{-1.72}$ & $2.36$\\
    \hline
    $\chi^{2}_{min}$ & $173.36$ & $17.71$ & $202.28$ & - \\
    \hline
    $\chi^{2}_{red}$ & $1.82$ & $0.66$ & $1.61$ & - \\
    \hline
    \end{tabular}
    \caption{ Mean values for the cosmographic parameters for the 4th order expansion, obtained from SLS ($y_s\leq0.6$), OHD and joint analysis ($y_s\leq0.6$), using the y-redshift parameterization and their uncertainties at $1\sigma$. The $\Lambda$CDM values obtained from \citet{Dunsby:2015ers} with $\Omega_{0m}=0.311$ \citep{Aghanim:2018} are presented for comparison.}
    \label{SLSandCronomvalues}
\end{table}

Finally, we present the analysis of the effective EoS constructed from the cosmographic parameters at fourth order. We consider that there are two important epochs to explore the effective EoS: the current Universe ($z=0$) and the transition phase, where the Universe evolves from a non-accelerated to an accelerated stage. Several observations indicate that this transition happens at $z\approx0.7$  \citep{Aghanim:2018} or, in terms of $y$-redshift, at $y\approx0.4$. We present the EoS at $y=0.4$ ($z=0.7$) and $y=z=0$ for the SLS, OHD and the Joint analysis respectively in Table \ref{Omegasz}.
The values for the Joint and SLS analysis are in good agreement with the $\Lambda$CDM model, where the EoS tends to $\sim-0.689$ at $z=0$ (result obtained using values presented below Eq. \eqref{ll}). In addition, the effective equation of state at the transition epoch ($y=0.4$) for the Joint analysis is $w_{eff}\sim-0.295^{+0.046}_{-0.052}$ (see Table \ref{Omegasz}), which is in good agreement with the value of EoS in $\Lambda$CDM ($w_{eff}\sim-0.310$).

\begin{figure}
    \includegraphics[width=0.47\textwidth]{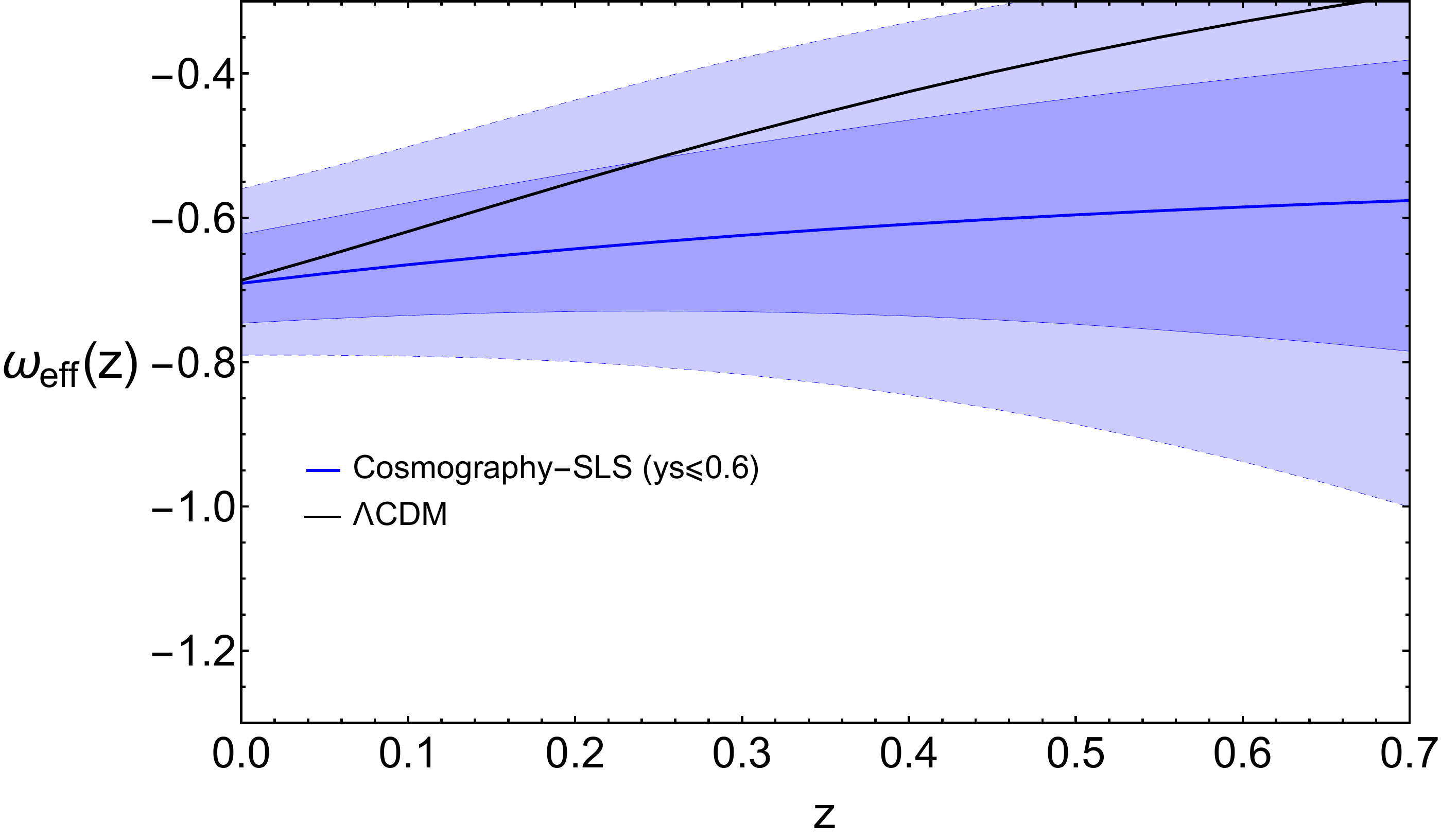}
     \includegraphics[width=0.47\textwidth]{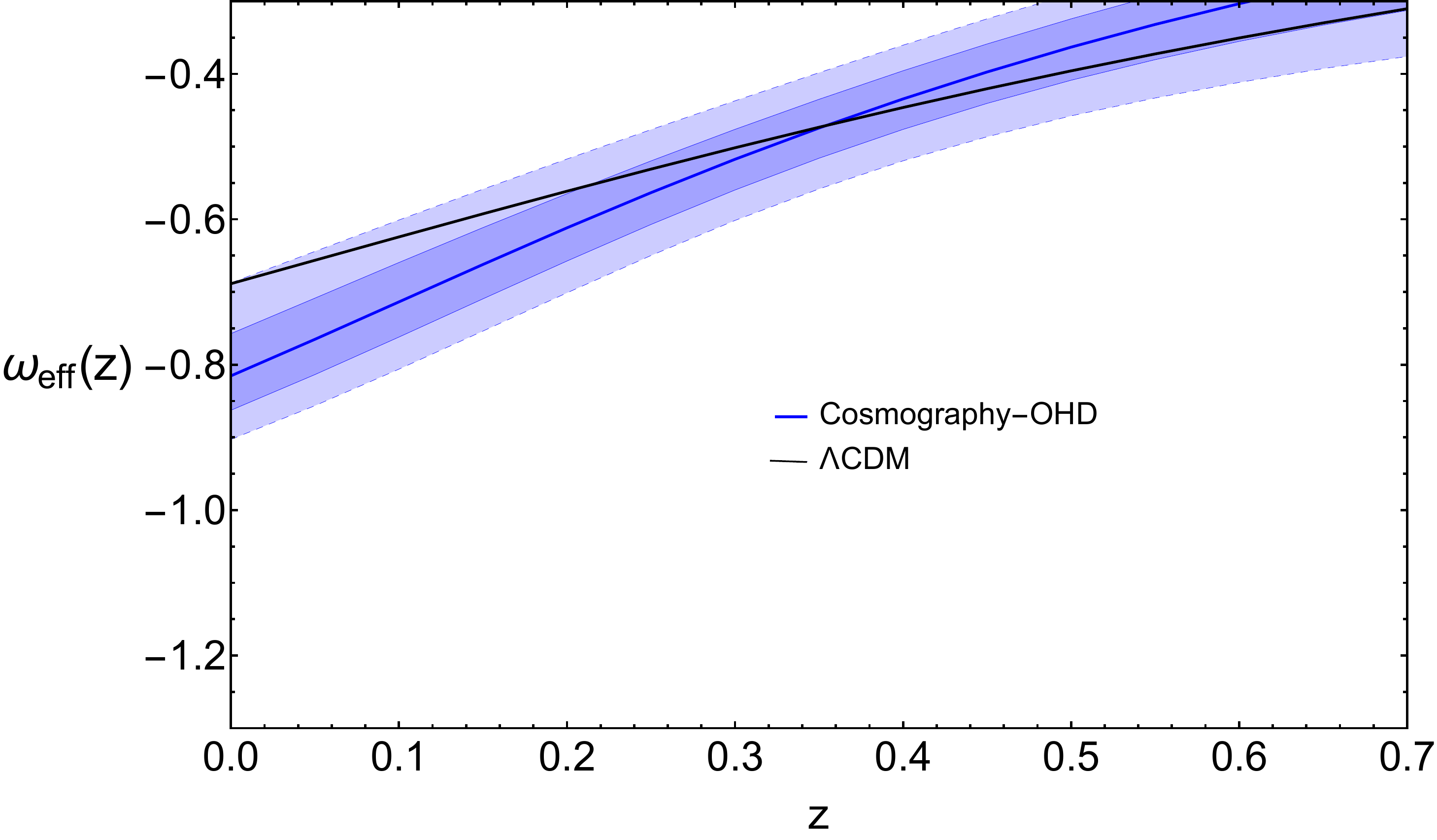}
     \includegraphics[width=0.47\textwidth]{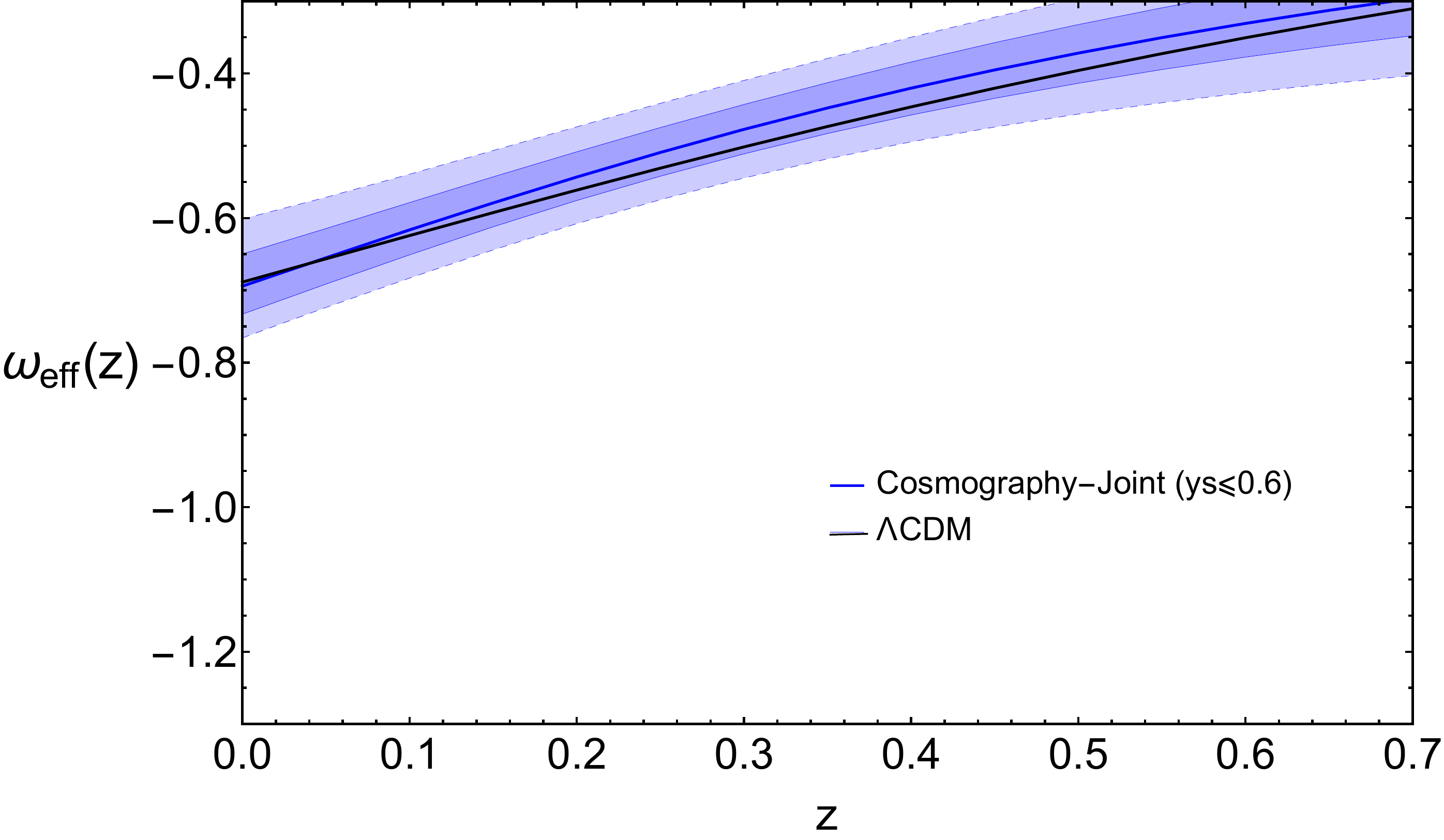}
    \caption{Reconstruction of $w_{eff}$ from the SLS (Top), OHD (Middle) and Joint (Bottom) constraints.For Comparison, the effective EoS for $\Lambda$CDM is also showed (black solid line). In the case of the standard cosmological model, we use the cosmographic parameters of Eqs. \eqref{qq}-\eqref{ll} under the assumption of $\Omega_{0m}$ as reported by \citet{Aghanim:2018} and the Taylor series of cosmography. The EoS is written in terms of the redshift through the parameterization $y=z/(1+z)$. The uncertainty bands correspond to $1$ and $2 \sigma$ of the SLS reconstruction.}
    \label{EoSSLSOHDJOINT}
\end{figure}

Figure \ref{EoSSLSOHDJOINT}, shows the effective EoS reconstruction in terms of redshift in the interval $0<z<0.7$ for the SLS (top panel), OHD (middle panel) and Joint (bottom panel) constraints together with the effective EoS reconstruction for the $\Lambda$CDM cosmographic parameters. They provide a graphic confirmation that the EoS predicted by cosmography is in concordance with $\Lambda$CDM at $z=0$ for the SLS and Joint cases, while showing a difference with the standard model for the OHD case, because at $z=0$, $w_{eff}=-0.815$. On the other hand, at large redshifts the discrepancy between cosmographic results for SLS sample and the standard model is important, for instance, there are remarkable differences in the epoch of transition ($z\approx0.7$). Notice that the EoS in the case of SLS reconstruction (Fig. \ref{EoSSLSOHDJOINT}, top) is a slightly decreasing function, while in the case of OHD and Joint (Figs. \ref{EoSSLSOHDJOINT}, middle and bottom) are increasing functions in the interval $0<z<0.7$. These differences can be due to the fact that cosmological parameters are estimated from two different expressions, i.e.  from Eq. \eqref{Hubbley} and Eq. \eqref{Dy} for OHD and strong lensing measurements respectively, obtaining a larger scatter for SLS in comparison with $\Lambda$CDM. Other possible reason, could be that the OHD and SLS redshift are different intervals, generating that the approximation does not coincide with the theoretical result. Regarding the results for EoS given by \cite{Rezaei:2020lfy}, they obtain that at $z=0$ the Universe is dominated by a phantom phase ($w<-1$), while in our case it points towards a quintessence phase ($-1<w<-1/3$). In this case, the Universe will not have a future singularity (Big Rip).

\begin{table}
    \centering
    \begin{tabular}{|c|c|c|}
    \hline
        Equation of State &  y=0.4 (z=0.7) & z=y=0 \\
    \hline
        \multicolumn{3}{|c|}{SLS ($y_s\leq0.6$)}\\
    \hline
        $w_{eff}(y)$  & $-0.580^{+0.191}_{-0.209}$ & $-0.690^{+0.068}_{-0.056}$ \\
    \hline    
        \multicolumn{3}{|c|}{OHD}\\
    \hline
        $w_{eff}(y)$ & $-0.254^{+0.043}_{-0.058}$ & $-0.815^{+0.058}_{-0.047}$ \\
    \hline
    \multicolumn{3}{|c|}{Joint ($y_s\leq0.6$)}\\
    \hline
        $w_{eff}(y)$ & $-0.296^{+0.047}_{-0.052}$ & $-0.694^{+0.045}_{-0.039}$ \\
    \hline    
    \end{tabular}
    \caption{Effective equation of state expanded at 4th order, explored in the transition epoch $y=0.4$ ($z=0.7$) and in the current Universe $y=z=0$, using SLS ($y_s\leq0.6$), OHD, and Joint ($y_s\leq0.6$).}
    \label{Omegasz}
\end{table}

\section{Conclusions and Outlooks} \label{Con}

Cosmography has been one of the most important ways to find the dynamics of our Universe through the assistance of modern observations. Consistency in the cosmographic parameter estimation is found by including different sets of data and techniques.

Strong lensing systems has been used in the constriction of the free parameters of several cosmological models, it is a complementary data-set to test diverse DE models as those studied by \cite{Amante:2019xao}. In this work we used, for the first time, the SLS data to constraint the cosmographic parameters. For this task, we constructed the angular diameter distance ratio $D_{obs}$ (given by Eq. \eqref{Dlens}) in terms of the cosmographic parameters through the expansion in Taylor series of the angular diameter distances for the SLS and the $H(z)$ function. An MCMC analysis was used to get the best fit parameters with the sample of SLS observations and OHD compiled by \citet{Amante:2019xao} and \citet{Magana:2017nfs}, respectively.

As mentioned before, this is the first time when a $D^{th}$ function is used in a cosmographic analysis through SLS data (see Eqs. \eqref{dthz} and \eqref{dthy}). We clearly observe that the $y$-redshift approach has a decay in the $y_{s}>0.6$ region for the angular diameter distance, which is evidence of the deviation of the $y$-redshift parameterization, maybe due to the lack of terms in our Taylor series expansion or the evaluation of the series in one of the the boundaries of the interval of expansion. This could be the reason for the large difference in the parameters at high redshift ($z > 1.5$, $y>0.6$) (see results in Appendix \ref{AP1}). Although the $y$-parameterization is not ideal, we obtain better constraints than when using the $z$-redshift by itself: the Taylor series will not converge for objects with $z>1$. Furthermore, the use of the $y$-redshift parameterization coupled with the restriction $y_s\leq0.6$ ameliorates both the discrepancies produced by expanding a Taylor series in one of the boundaries of the interval and the nonphysical values (i.e. $D^{th}>1$) in the theoretical lens equation \eqref{einstrad}. 

The estimation of the cosmographic parameters with a fourth order approximation are summarized in Table \ref{SLSandCronomvalues}. The value of the jerk parameter is particularly interesting because it indicates whether the responsible for the universe acceleration is a constant ($j=1$) or a dynamical dark energy ($j\neq1$). Although studies \citep[see][for instance]{Zhang:2016urt}
with SNeIa samples fail to reproduce the theoretical value $j_0=1$, it is important to highlight that the SLS sample is in agreement with the theory. The jerk parameter constraint ($j_0=1.22^{+0.29}_{-0.34}$) slightly point towards a dynamical dark energy in contrast with the standard cosmological model\footnote{ Cosmographic analysis usually predicts a dynamical DE.}. A direct consequence of this result is that a dynamical DE could be an elegant solution to the $H_0$ tension among late and early observations \citep{Bernal:2016gxb}. In spite of this, we consider this result not conclusive because we are only analyzing up to a fourth order in an infinite series; i.e., other corrections together with other data (e.g. Cosmic Microwave Background Radiation, Baryon Acoustic Oscillations, among others) are necessary to estimate the value of the jerk parameter and elucidate whether we are dealing with a CC or a dynamical DE.

Finally, we reconstructed the effective EoS using SLS, OHD and Joint analysis compilations, and compare them with the standard EoS of $\Lambda$CDM. Our results for the cosmographic parameters yield an EoS with $-1<w<-1/3$, which is inside the region where, according to General Relativity, an effective fluid accelerates the cosmic expansion. We emphasize that, although the three samples (SLS, OHD and Joint) produce EoS that fall in the region for an accelerated Universe, the best fits at $z=0$ for the $\Lambda$CDM model are obtained with the SLS and the Joint analysis. However, at high redshifts there are important differences with $\Lambda$CDM, even at the transition epoch ($z\approx0.7$) from decelerating to accelerating expansion.

We conclude that SLS is in its first steps of being an efficient tool to be used in the cosmography method. We propose to increase the number of SLS data in order to improve the statistic and increase the order in the Taylor series. In addition, we also encourage the exploration of other parameterizations \cite[see for example][]{Aviles:2012ay}
employing SLS samples.

\section*{Acknowledgments}
We thank the anonymous referee for thoughtful remarks and suggestions. The authors acknowledge the enlightening conversation with Michel Cur\'e, Alejandro Aviles and Marek Demianski. A.L. and M.H.A. acknowledges support from CONACYT M. Sc. and PhD fellow respectively, Consejo Zacatecano de Ciencia, Tecnolog\'{\i}a e Innovaci\'on (COZCYT) and Centro de Astrof\'{\i}sica de Valpara\'{\i}so (CAV). A. L. thanks the staff of the Instituto de F\'{\i}sica y Astronom\'{\i}a of the Universidad de Valpara\'{\i}so where part of this work was done.
M.A.G.-A. acknowledges support from CONACYT research fellow, Sistema Nacional de Investigadores (SNI) and Instituto Avanzado de Cosmolog\'ia (IAC). J.M. acknowledges the support from CONICYT project Basal AFB-170002, V.M. acknowledges the support of  Centro de Astrof\'{\i}sica de Valpara\'{\i}so (CAV). V.M., J.M. and M.A.G.-A. acknowledge ANID REDES (Grant No. 190147).

\section*{Data Availability}
The data underlying this article were accessed from https://doi.org/10.1093/mnras/sty260, in the arXiv repository e-Print:1906.04107 and references therein.

\bibliographystyle{mnras}
\bibliography{librero0}

\appendix
\section{Results using the full SLS sample} \label{AP1}

\begin{figure}
    \includegraphics[width=0.4\textwidth]{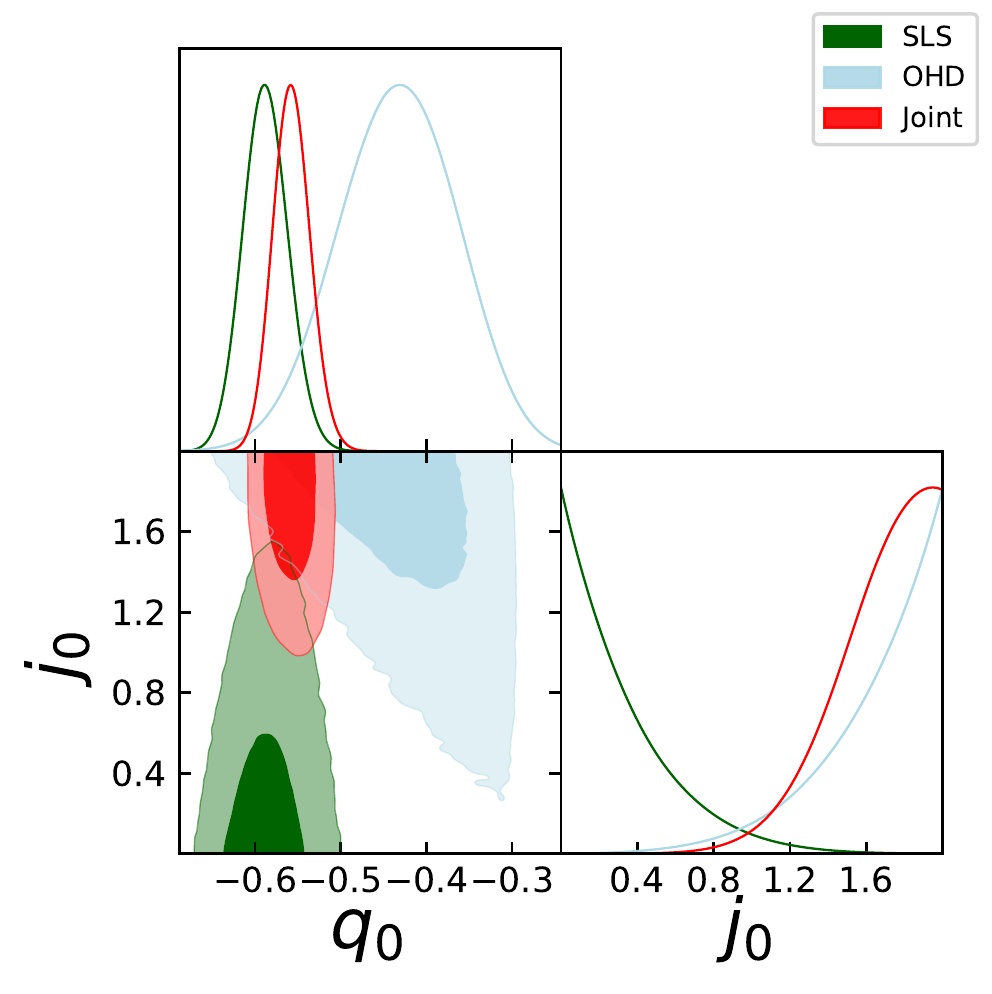}
    \caption{1D marginalized posterior distributions and the 2D $68\%$, $99.7\%$ confidence levels for $q_0$, and $j_0$ parameters from SLS, OHD and a joint analysis. The series were truncated up to the second order.}
    \label{q0j0App}
\end{figure}

\begin{figure}
    \includegraphics[width=0.5\textwidth]{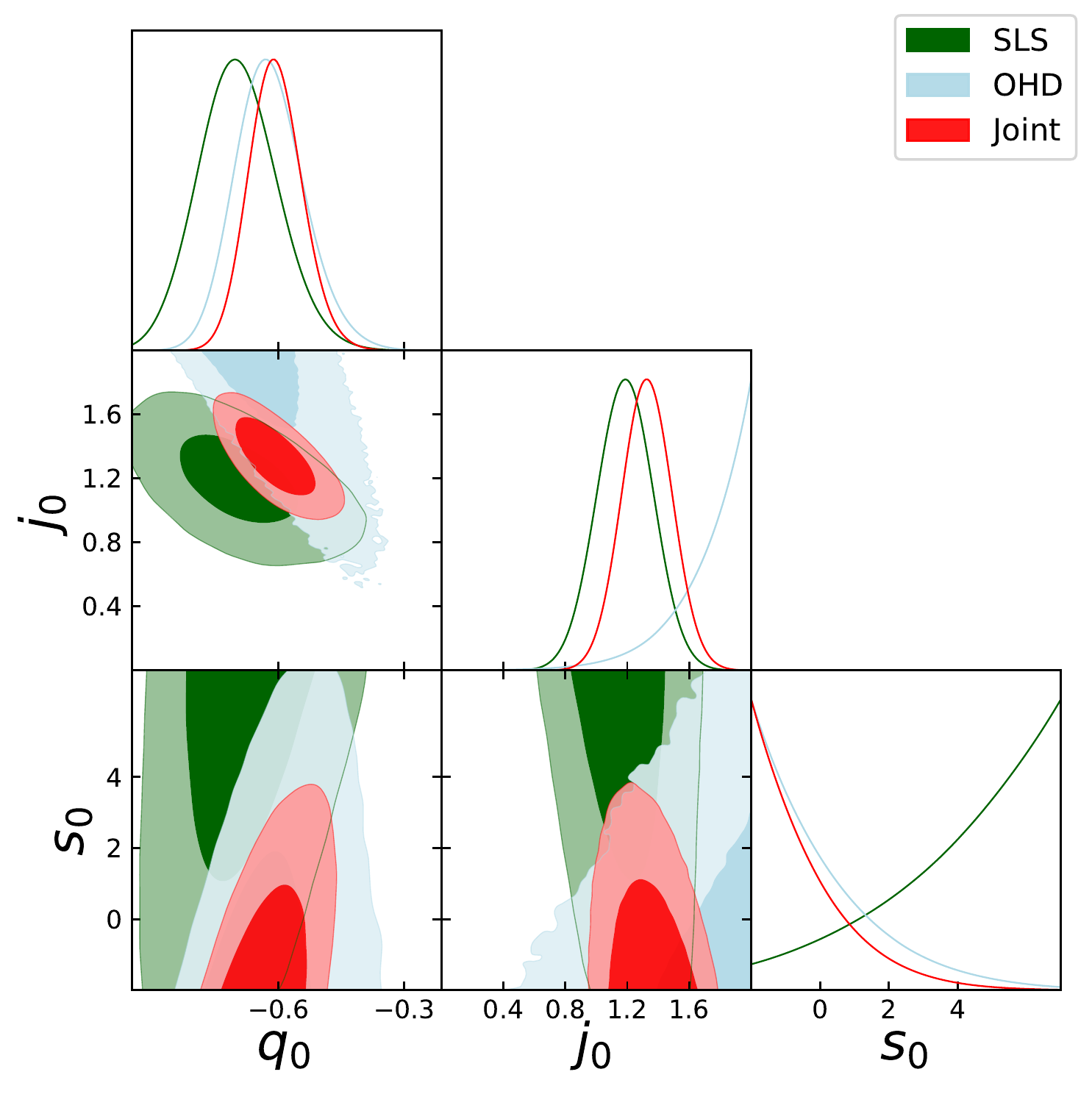}
    \caption{1D marginalized posterior distributions and the 2D $68\%$, $99.7\%$ confidence levels for $q_0$, $j_0$, and $s_0$ parameters from SLS, OHD and a Joint analysis. The series were truncated up to the third order.}
    \label{q0j0s0App}
\end{figure}

\begin{figure}
    \includegraphics[width=0.5\textwidth]{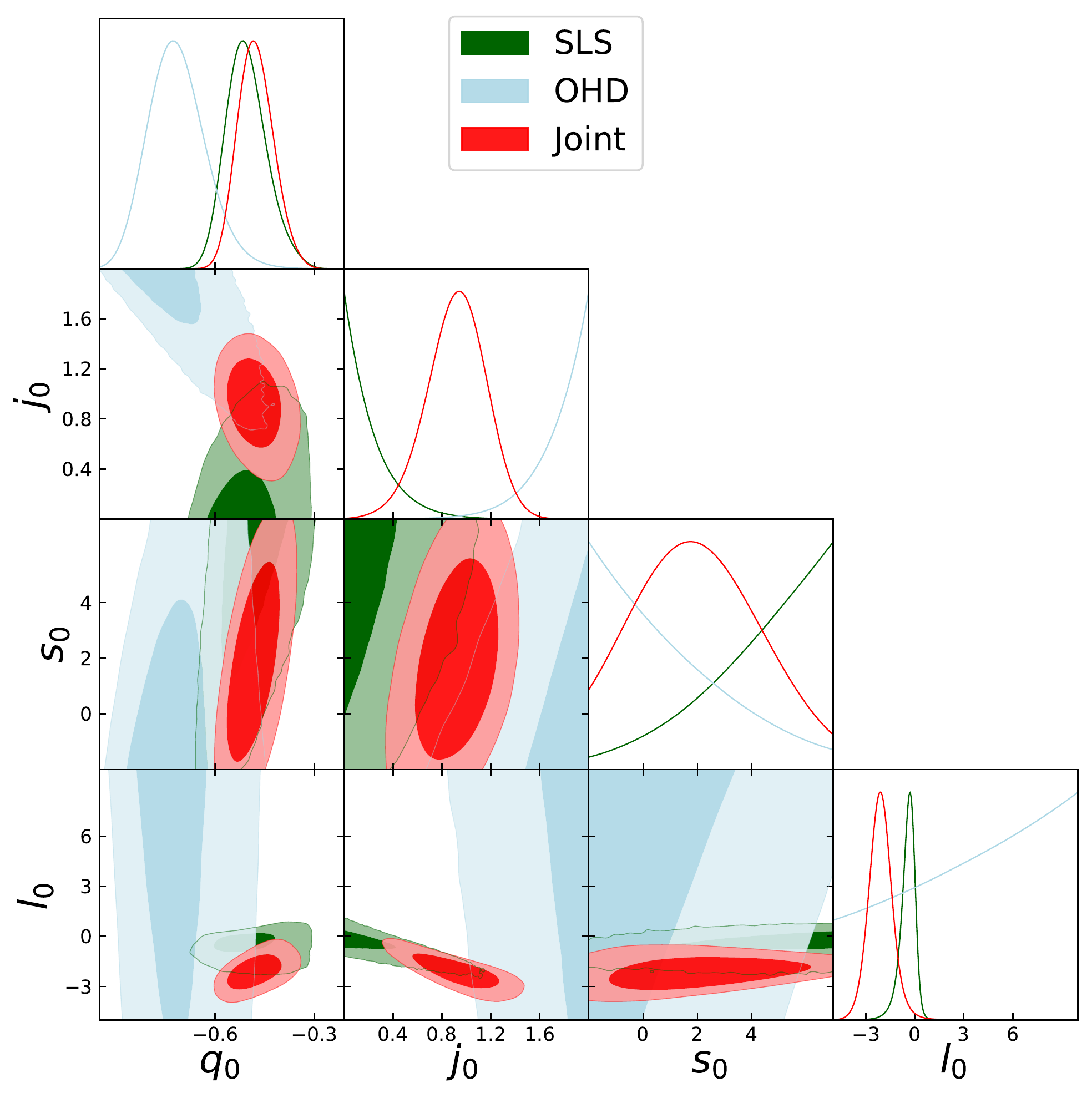}
    \caption{1D marginalized posterior distributions and the 2D $68\%$, $99.7\%$ confidence levels for $q_0$, $j_0$, $s_0$ and $l_0$ parameters from SLS, OHD and a Joint analysis. The series were truncated up to the fourth order.}
    \label{q0j0s0l0App}
\end{figure}

\begin{figure}
    \includegraphics[width=0.47\textwidth]{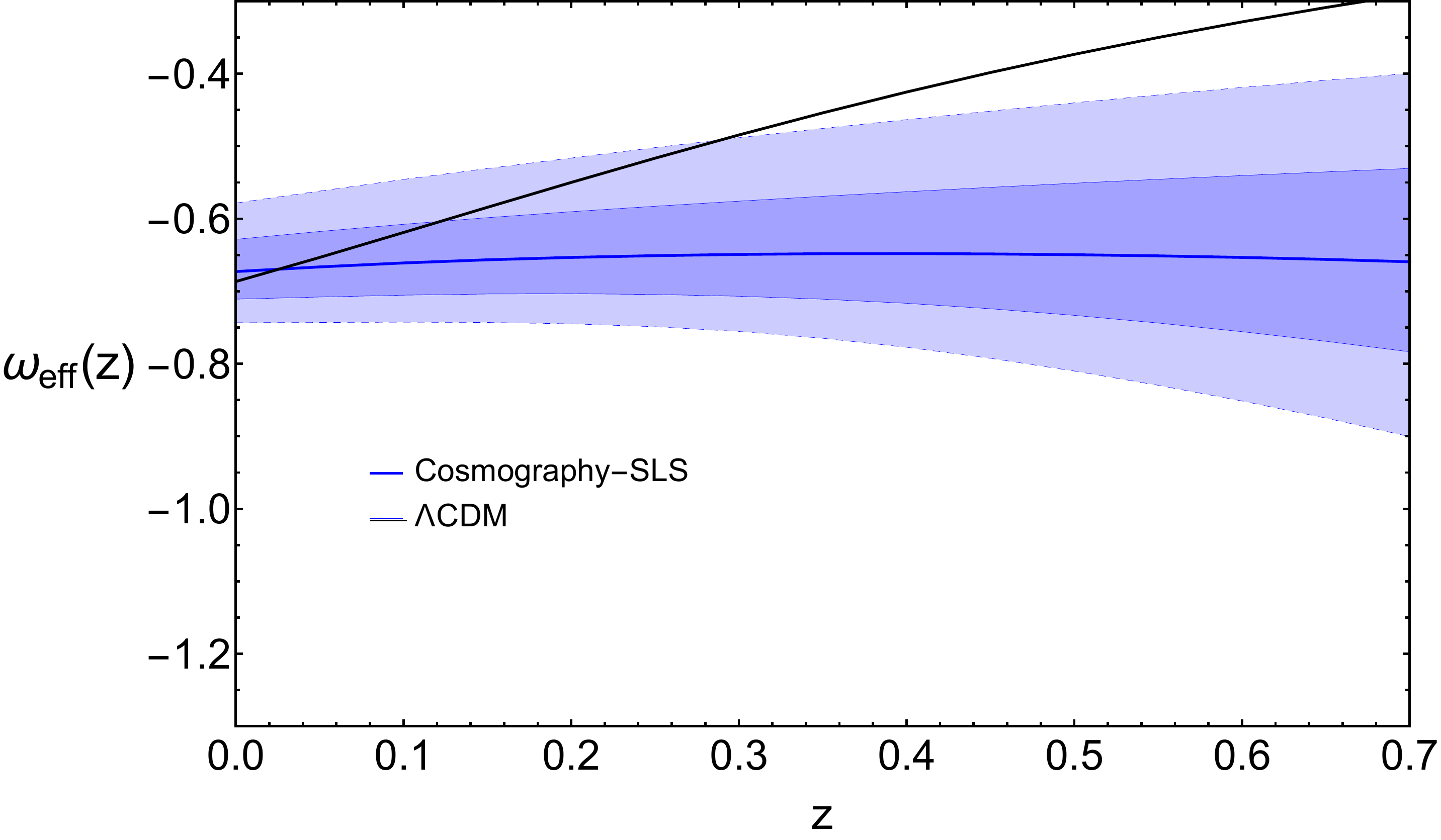}
     \includegraphics[width=0.47\textwidth]{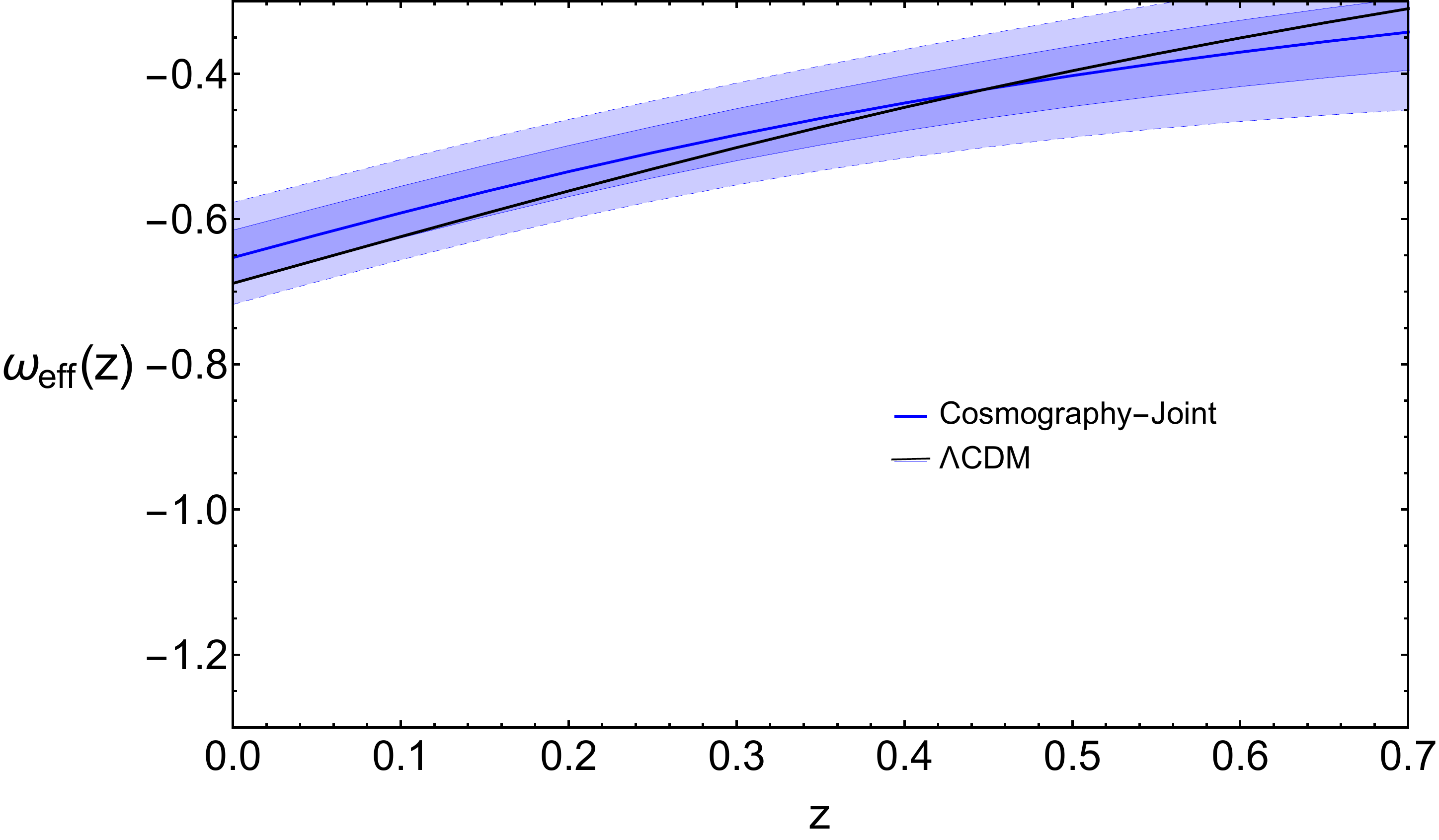}
    \caption{Reconstruction of $w_{eff}$ from the SLS (Top) and Joint (Bottom) constraints. For comparison, the effective EoS for $\Lambda$CDM  is also showed (black solid line). In the case of the standard cosmological model, we use the cosmographic parameters of Eqs. \eqref{qq}-\eqref{ll} under the assumptions of $\Omega_{0m}$ as reported by \citet{Aghanim:2018} and the Taylor series of cosmography. The EoS is written in terms of the z-redshift through the parameterization $y=z/(1+z)$. The uncertainty bands correspond to $1$ and $2 \sigma$ of the SLS reconstruction.}
    \label{EoSSLSAPP}
\end{figure}

In this Appendix we present the results obtained using the full SLS sample, i.e. without the constriction $y_s<0.6$, in order to complement those shown inside the analysis presented in Section \ref{Res}. Our results are summarized as follows: In Table \ref{SLSandCronomvaluesApp} are presented the cosmographic parameters at second, third and fourth order, being complemented with Figs. \ref{q0j0App}-\ref{q0j0s0l0App}, showing the 1D marginalized posterior distributions and the 2D at $68\%$, $99.7\%$ confidence levels for the cosmographic parameters. Furthermore, Table \ref{OmegaszApp} contains the effective EoS at $y=0.4$ $(z=0.7)$ and $z=y=0$ for the full SLS and Joint samples \footnote{{The table does not contains the $w_{eff}(y)$ for OHD because it is the same result shown in Table \ref{Omegasz}.}}, Figs. \ref{EoSSLSAPP}, Top and Bottom, describes the reconstruction of the effective EoS under the SLS and Joint data with the full sample and its respective comparison with $\Lambda$CDM model.

\begin{table}
    \centering
    \begin{tabular}{|c|c|c|c|}
    \hline
        Param. & 2nd order & 3rd order & 4th order \\
    \hline
        \multicolumn{4}{|c|}{143 SLS}\\
    \hline
        $q_0$ & $-0.59\pm0.07$ & $-0.68\pm0.09$ & $-0.51\pm0.06$  \\
    \hline
        $j_0$ & $0.24\pm0.35$ & $1.16\pm0.19$ & $0.16\pm0.24$ \\
    \hline
        $s_0$ & --- & $7.03\pm2.17$ & $4.69\pm2.5$ \\
    \hline
        $l_0$ & --- & --- & $-0.37\pm0.45$ \\
    \hline
    $\chi^{2}_{min}$ & 234.42 & 225.25 & 225.14 \\
    \hline
    $\chi^{2}_{red}$ & 1.66 & 1.61 & 1.62 \\
    \hline
        \multicolumn{4}{|c|}{31 OHD}\\
    \hline
        $q_0$ & $-0.43^{+0.07}_{-0.07}$ & $-0.63^{+0.08}_{-0.07}$ & $-0.72^{+0.09}_{-0.07}$ \\
    \hline
        $j_0$ & $1.73^{+0.20}_{-0.39}$ & $1.81^{+0.14}_{-0.30}$ & $1.82^{+0.13}_{-0.27}$ \\
    \hline
        $s_0$ & --- & $-0.51^{+2.23}_{-1.11}$ & $0.44^{+2.98}_{-1.77}$ \\
    \hline
        $l_0$ & --- & --- & $4.00^{+4.28}_{-5.61}$ \\
    \hline
    $\chi^{2}_{min}$ & $36.86$ & $23.50$ & $17.71$ \\
    \hline
    $\chi^{2}_{red}$ & $1.27$ & $0.84$ & $0.66$ \\
    \hline
  \multicolumn{4}{|c|}{Joint (143 SLS+ 31 OHD)}\\
    \hline
        $q_0$ & $-0.56^{+0.02}_{-0.02}$ & $-0.61^{+0.06}_{-0.06}$ & $-0.48^{+0.06}_{-0.05}$ \\
    \hline
        $j_0$ & $1.70^{+0.20}_{-0.30}$ & $1.33^{+0.16}_{-0.16}$ & $0.93^{+0.23}_{-0.24}$ \\
    \hline
        $s_0$ & --- & $-0.78^{+1.76}_{-0.90}$ & $1.96^{+2.37}_{-2.18}$ \\
    \hline
        $l_0$ & --- & --- & $-2.11^{+0.69}_{-0.68}$ \\
    \hline
    $\chi^{2}_{min}$ & $286.64$ & $259.20$ & $263.26$ \\
    \hline
    $\chi^{2}_{red}$ & $1.67$ & $1.52$ & $1.55$ \\
    \hline
    \end{tabular}
    \caption{ Mean values for the cosmographic parameters obtained from SLS, OHD, joint analysis, using the $y$-redshift parameterization and their uncertainties at $1\sigma$.}
    \label{SLSandCronomvaluesApp}
\end{table}

\begin{table}
    \centering
    \begin{tabular}{|c|c|c|}
    \hline
        Equation of State &  y=0.4 (z=0.7) & z=y=0 \\
    \hline
        \multicolumn{3}{|c|}{SLS}\\
    \hline
        $w_{eff}(y)$  & $-0.661^{+0.127}_{-0.124}$ & $-0.672^{+0.044}_{-0.038}$ \\
    \hline
  \multicolumn{3}{|c|}{Joint (SLS+OHD)}\\
    \hline
        $w_{eff}(y)$ & $-0.342^{+0.048}_{-0.052}$ & $-0.653^{+0.039}_{-0.034}$ \\
    \hline
    \end{tabular}
    \caption{Equation of state expanded at 4th order, explored in the transition epoch $y=0.4$ ($z=0.7$) and in the current Universe $y=z=0$, using SLS, OHD, and Joint.}
    \label{OmegaszApp}
\end{table}

{ 
\section{Consequences of the inclusion of a $f$ factor in the estimation of cosmographic parameters}\label{AP2}

As we mention before, SLS can be affected by different systematics, hence the lens galaxy mass distribution is not exactly isothermal, yielding more complicated profiles. To take into account these systematics, a corrective $f$ parameter can be introduced in $D^{obs}$ (Eq. \ref{Dlens}) as
\begin{equation}
D^{obs}\equiv\frac{D_{ls}}{D_{s}} = \frac{c^2 \theta_{E}}{4 \pi \sigma^2 f}. \label{Dlensf}
\end{equation} 
}

\begin{figure}
    \centering
    \includegraphics[width=0.47\textwidth]{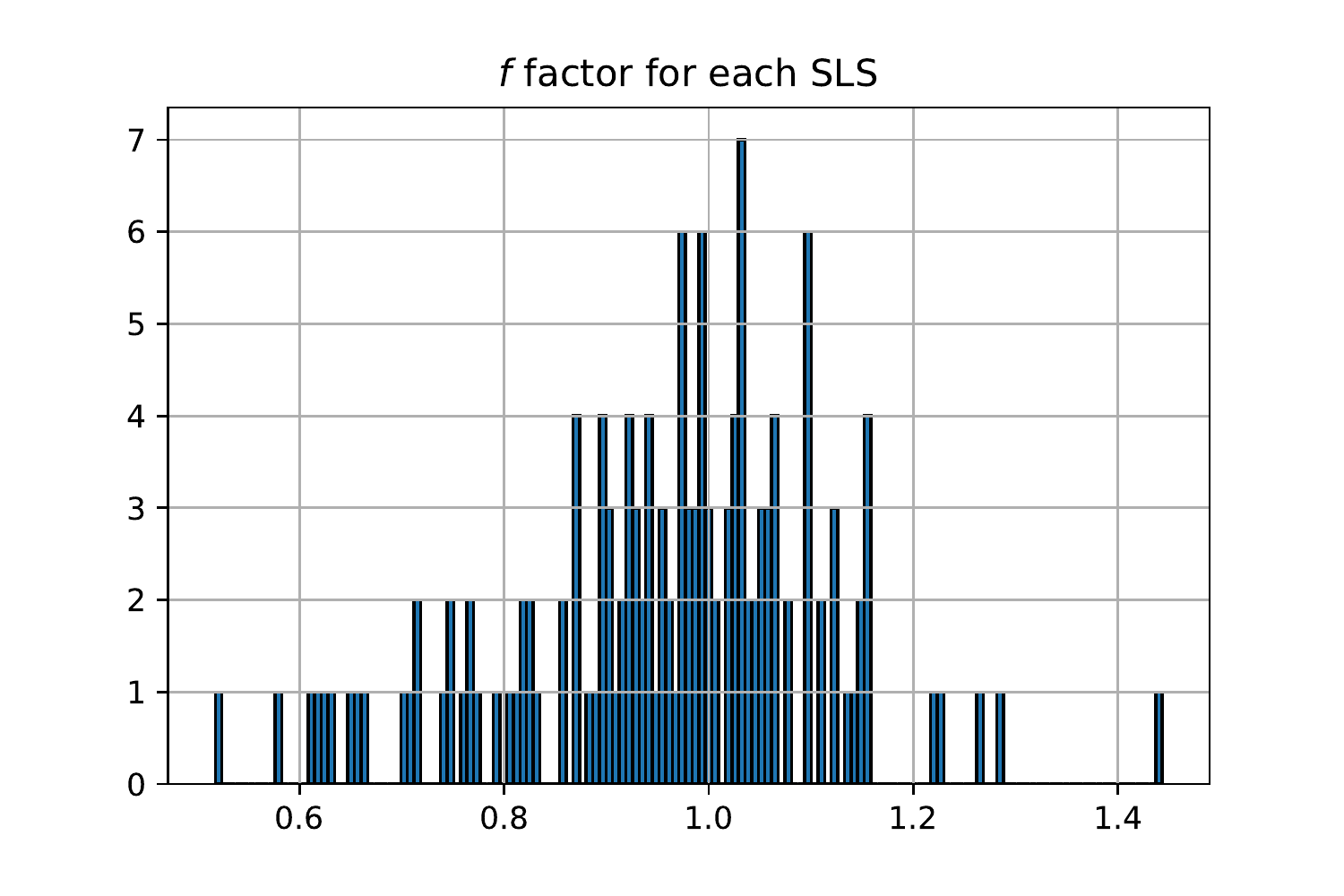}
    \caption{Histogram for the best fit values of the $f$ factor in each of the SLS for the full sample of 143 systems.}
    \label{histf}
\end{figure}

The inclusion of this parameter $f$ also reckons possible deviations of the SLS mass profile proposed. In order to take this into account, we implement two different tests for the estimation of cosmographic parameters: i) we use $f$ as an individual parameter for each SLS, and ii) we assume a single value of $f$ for all the SLS data. The first analysis was carried out optimizing the parameters trough the differential evolution
method from {\it scipy} python package. We obtain values closer to the isothermal profile in the majority of the systems as shown in figure \ref{histf}, with an average value of $f =0.96$ and average standard deviation of $0.15$ for the full sample of 143 SLS. A factor $f$ close to unity means no relevant variations from the original proposal. 
In the second case, we found that the values for the cosmographic parameters are consistent, within 1 $\sigma$ of confidence level, with the previous analysis. These results are presented in Tables \ref{factorfvalues}, \ref{factorfvalues2}.

\begin{table}
    \centering
    \begin{tabular}{|c|c|c|c|}
    \hline
        Param. & 143 SLS & 143 SLS with $f$ & $\Lambda$CDM  \\
    \hline
    \multicolumn{4}{|c|}{}\\
    \hline
        $q_0$ & $-0.51\pm0.06$ & $-0.45\pm0.09$ &  $-0.53$\\
    \hline
        $j_0$ & $0.16\pm0.24$ & $0.15\pm0.23$ &  $1$ \\
    \hline
        $s_0$ & $4.69\pm2.5$ & $4.79\pm2.6$ &  $-0.39$\\
    \hline
        $l_0$ & $-0.37\pm0.45$ & $-0.38\pm0.37$ &  $2.36$\\
    \hline
        $f$ & - & $1.008\pm0.01$ & - \\
    \hline
    $\chi^{2}_{min}$ & $225.14$ & $224.91$ &  - \\
    \hline
    $\chi^{2}_{red}$ & $1.62$ & $1.62$ &  - \\
    \hline
    \end{tabular}
    \caption{Mean values for the cosmographic parameters for the 4th order expansion, obtained from SLS in two cases: without the corrective factor $f$ and with the corrective factor $f$ mentioned in Eq.\eqref{Dlensf}. The $\Lambda$CDM values obtained from \citet{Dunsby:2015ers} with $\Omega_{0m}=0.311$ \citep{Aghanim:2018} are presented for comparison.}
    \label{factorfvalues}
\end{table}

\begin{table}
    \centering
    \begin{tabular}{|c|c|c|c|}
    \hline
        Param. & 99 SLS & 99 SLS with $f$ & $\Lambda$CDM  \\
    \hline
    \multicolumn{4}{|c|}{}\\
    \hline
        $q_0$ & $-0.54\pm0.10$ & $-0.51\pm0.15$ &  $-0.53$\\
    \hline
        $j_0$ & $0.35\pm0.52$ & $0.33\pm0.48$ &  $1$ \\
    \hline
        $s_0$ & $3.7\pm3.3$ & $3.66\pm3.2$ &  $-0.39$\\
    \hline
        $l_0$ & $-0.35\pm2.2$ & $-0.43\pm2.3$ &  $2.36$\\
    \hline
        $f$ & - & $1.002\pm0.01$ & - \\
    \hline
    $\chi^{2}_{min}$ & $173.36$ & $173.4$ &  - \\
    \hline
    $\chi^{2}_{red}$ & $1.82$ & $1.84$ &  - \\
    \hline
    \end{tabular}
    \caption{Mean values for the cosmographic parameters for the 4th order expansion, obtained from SLS data, without the SLS with $y_s \leq 0.6$, in two cases: without the corrective factor $f$ and with the corrective factor $f$ mentioned in Eq.\eqref{Dlensf}. The $\Lambda$CDM values obtained from \citet{Dunsby:2015ers} with $\Omega_{0m}=0.311$ \citep{Aghanim:2018} are presented for comparison.}
    \label{factorfvalues2}
\end{table}

\begin{table}
    \centering
    \begin{tabular}{|c|c|c|c|}
    \hline
        Param. & Joint & Joint with $f$ & $\Lambda$CDM  \\
    \hline
    \multicolumn{4}{|c|}{}\\
    \hline
        $q_0$ & $-0.54\pm0.10$ & $-0.44\pm0.07$ &  $-0.53$\\
    \hline
        $j_0$ & $1.22\pm0.34$ & $0.77\pm0.3$ &  $1$ \\
    \hline
        $s_0$ & $0.86\pm3.1$ & $1.62\pm2.4$ &  $-0.39$\\
    \hline
        $l_0$ & $-2.31\pm2.7$ & $-1.86\pm0.9$ &  $2.36$\\
    \hline
        $f$ & - & $1.007\pm0.01$ & - \\
    \hline
    $\chi^{2}_{min}$ & $202.28$ & $263.2$ &  - \\
    \hline
    $\chi^{2}_{red}$ & $1.61$ & $1.65$ &  - \\
    \hline
    \end{tabular}
    \caption{Mean values for the cosmographic parameters for the 4th order expansion, obtained from SLS data, without the SLS with $y_s \leq 0.6$ together with the data of the OHD systems, we show here the two cases: without the corrective factor $f$ and with the corrective factor $f$ mentioned in Eq.\eqref{Dlensf}. The $\Lambda$CDM values obtained from \citet{Dunsby:2015ers} with $\Omega_{0m}=0.311$ \citep{Aghanim:2018} are presented for comparison.}
    \label{factorfvalues3}
\end{table}

\section{Mock data with SLS systems}\label{AP3}

We created a mock data set from the original SLS sample to investigate whether we can recover the cosmographic parameters used or not.
To simulate our data, we assume that $z_S$, $D_{ls}$, and $\sigma$ are distributed using the ranges provided by the SLS sample. 
To generate a lens system we proceed as follows: we take random values ($z_s$, $D_{ls}$, $\sigma$) from the observed distribution with an estimated random error of 10\% to 20\%. From the above, we estimate $z_l$ according to a $\Lambda$CDM cosmology.
Later, we choose random values for other parameters of the lens model (axis ratio $b/a=0.7\pm0.16$, shear $\gamma:[0.03,0.07]$, position angles $[0.,180.]\deg$) according to the work provided by \cite{Oguri_2010} and a power law exponent of the mass distribution $2.078\pm0.027$ obtained from \cite{Auger_2010} (with the value 2 yielding the SIE profile). Finally, choosing the position of the source, we modeled each individual system using \texttt{lensmodel} \citep{Keeton:2001ss} to obtain the Einstein radius and the position of the images. From the full sample of the mock data, we selected those systems that were modeled as SIS with $\Omega_0 = 0.27$, $H_0 = 0.8$, and with the condition proposed $y_s \leq 0.6$. We also considered the region of $0.5 \leq D^{obs} \leq 1.0$ because, as shown by the full sample of SLS data and argued in \cite{Amante:2019xao}, the objects in the region $D^{obs}\leq0.5$ cannot be modeled properly, yielding values for the $\chi^2$ far from the expected ($\chi^2\geq5$). Applying these conditions, we obtain a sample of 42 SLS from the mock data. The values obtained for the cosmographic parameters are shown in Table \ref{mockparam}. It is worth to mention that we were able to recover the input mock values for the cosmographic parameters at $1\sigma$ level of confidence, this is evidence of the accuracy of this method in the estimation of the cosmographic parameters.

\begin{table}   
    \centering
    \begin{tabular}{|c|c|c|}
    \hline
    Parameter & 42 Mock & 99 SLS \\ \hline
    \multicolumn{3}{|c|}{}\\ \hline
    $q_0$ & $-0.45\pm0.18$ & $-0.54\pm0.1$ \\ \hline
    $j_0$ & $0.74\pm0.75$ & $0.35\pm0.52$ \\ \hline
    $s_0$ & $2.45\pm3.08$ & $3.7\pm3.3$ \\ \hline
    $l_0$ & $1.39\pm5.5$ & $-0.35\pm2.2$ \\ \hline
    \multicolumn{3}{|c|}{}\\ \hline
    $\chi^2_{min}$ & $15.85$ & $173.36$\\ \hline
    $\chi^2_{red}$ & $0.417$ & $1.82$\\ \hline
    \end{tabular}
    \caption{Values for the cosmographic parameters obtained from the simulated mock SLS data. The values obtained from sample of SLS with $y_s \leq 0.6$ are presented here for comparison.}
    \label{mockparam}
\end{table}

The values of the cosmographic parameters at $4th$ order were obtained with the \texttt{emcee} Python code \citep{Foreman_Mackey_2013}, 5000 steps on the MCMC, 1000 steps for the burn-in phase and 1000 walkers. Throughout the analysis we consider $f = 1$ as we did in the section of the SLS data  analysis. These values for the cosmographic parameters obtained from the mock data are consistent at $1\sigma$ level with the values obtained from both the full sample and the one with the $y_s \leq 0.6$ condition, and also with the values expected from the $\Lambda$CDM model.

\end{document}